\documentclass[preprint,12pt,authoryear]{elsarticle}

\usepackage{amssymb}

\journal{Nuclear Physics A}

\begin{document}

\begin{frontmatter}


\title{Supernova equations of state  including full nuclear ensemble  with in-medium effects}

\author[fias,cfca]{Shun Furusawa}
\author[numazu]{Kohsuke Sumiyoshi}
\author[waseda]{Shoichi Yamada}
\author[rikadai]{Hideyuki Suzuki}
\address[fias]{Frankfurt Institute for Advanced Studies, J.W. Goethe University, 60438 Frankfurt am Main, Germany}
\address[cfca]{Center for Computational Astrophysics, National Astronomical Observatory of Japan, Osawa, Mitaka, Tokyo, 181-8588, Japan}
\address[numazu]{Numazu College of Technology, Ooka 3600, Numazu, Shizuoka 410-8501, Japan}
\address[waseda]{Department of Science and Engineering,
 Waseda University, 3-4-1 Okubo, Shinjuku, Tokyo 169-8555, Japan}
\address[rikadai]{Faculty of Science and Technology, Tokyo University of Science, Yamazaki 2641, Noda, Chiba 278-8510, Japan}
\begin{abstract}
We construct new equations of state for baryons at sub-nuclear densities for the use in core-collapse supernova simulations.
The abundance of various nuclei is obtained together with thermodynamic quantities. 
The formulation is an extension of the previous model,
 in which we adopted
the relativistic mean field theory with the TM1 parameter set 
for nucleons,
 the quantum approach for $d$, $t$, $h$ and  $\alpha$  as well as   the liquid drop model for the other nuclei under the nuclear statistical equilibrium. 
 We reformulate the model of the light nuclei other than $d$, $t$, $h$ and  $\alpha$
 based on the  quasi-particle description. 
Furthermore, we modify the  model
 so that the temperature dependences of surface and shell energies of  heavy nuclei could be taken into account.
The pasta phases for heavy nuclei and  the Pauli- and self-energy shifts  for $d$, $t$, $h$ and  $\alpha$ are taken into account in the same way as in the previous model.
We find that nuclear composition is considerably affected by the modifications in this work,  whereas thermodynamical quantities are not changed much.  In particular,  the washout of shell effect has a great impact on the mass distribution above  $T \sim 1$~MeV. This improvement may have an important effect on the rates of electron captures and coherent neutrino scatterings on nuclei in supernova cores.
\end{abstract}


\end{frontmatter}

\section{Introduction}
Hot and dense matter can be realized in  core collapse supernovae, which occur at the end of the evolution of massive stars and lead to
the formations of  a neutron star  or a black hole, the emissions of neutrinos and  gravitational waves and the synthesis of heavy elements.
The mechanism of this event is not clearly understood yet
because of their intricacies (see e.g. \cite{janka12,kotake12,burrows13}).
One of the underlying problems in these events is the equations of state (EOS's) of hot and dense matter
 both at sub- and supra-nuclear densities. 
EOS provides information on compositions of nuclear matter in addition to thermodynamical quantities such as pressure, entropy and sound velocity.
The compositions  play important roles to determine the evolution of the lepton fraction through weak interactions.
This lepton fraction is  one of the most critical ingredients for the dynamics of core collapse supernovae \citep{hix03,lentz12}. 

The EOS for the simulations of core collapse supernovae must cover a wide range of density ($10^5 \lesssim \rho_B \lesssim 10^{15} \rm{g/cm^3}$) and 
temperature ($0.1 \lesssim T \lesssim 10^{2}$~MeV), including both neutron-rich and proton-rich regimes. 
One of the difficulties in constructing the EOS is originated from the fact that depending on the density, temperature and proton fraction,
 the matter consists of either dilute free nucleons or
 a mixture of nuclei and free nucleons or strongly interacting dense nucleons.
Furthermore, there are uncertainties not only  in the
description of homogeneous matter
but also of the in-medium effect for nuclei  at finite densities and temperatures  \citep{aymard14,agrawal14a}. 
The fact makes the variation of supernova EOS's
\citep{steiner13,buyukcizmeci13}.
At high temperatures ($T \gtrsim 0.4 $~MeV), chemical equilibrium is achieved for all strong and electromagnetic reactions, 
which is referred to as nuclear statistical equilibrium and the nuclear composition is determined as a function of density,
 temperature, and proton fraction \citep{timmes99, blinnikov11}.
In this paper, we are concerned with the high temperature regime, in which the nuclear composition can be treated as a part of EOS.

There are two types of EOSs for supernova simulations.
The single nucleus approximation (SNA) is the one option, in which
 the ensemble of heavy nuclei
is represented by a single nucleus.
The other option is multi-nucleus EOS, in which thermal ensemble of various nuclei is solved
for each set up of thermodynamical condition.
Two  standard EOS's in wide use for the simulations of core-collapse  supernovae
 are  categorized to the former group; \cite{lattimer91} and   \cite{shen98a,shen98b,shen11} calculated  the  representative heavy nuclei by using the compressible liquid drop model   and Thomas Fermi model,  respectively. 
In such calculations, 
 in-medium effects on a single nucleus, such as compression and deformation, are taken into account more easily 
in comparison with the multi-nucleus EOS. 
On the other hand, they can never provide a nuclear composition, which is indispensable to estimate weak interaction rates in supernova simulations.
Furthermore,  even the average mass number and total mass fraction of heavy nuclei
may not be correctly reproduced by the representative nucleus due to the SNA approximation  \citep{burrows84,hempel10,furusawa11}.

In this decade, some multi-nucleus EOS's have been formulated by different research groups.
SMSM EOS \citep{botvina04, botvina10,buyukcizmeci14} is  a generalization of the statistical model
 of multi-fragmentation reactions induced by heavy-ion collisions \citep{bondorf95}.
In this model, temperature  dependences of bulk and surface energies are taken into account  based on the liquid drop model (LDM)
 although they ignored the shell effects of nuclei,
 which are important for reproducing the abundance of nuclei at low temperatures.
 \cite{hempel10} also construct EOS's (HS EOS)  
based on the relativistic mean field theory (RMF) with different parameters for nuclear interactions.
In this model, nuclear binding energies as well as shell effects are evaluated from experimental and theoretical mass data in vacuum.
They  ignored, on the other hand,  high-density and -temperature effects on nuclear bulk, surface  and shell energies, which are explained later.
The EOS  provided by G. Shen et al. \citep{sheng11}  is a hybrid model, 
where a multi-nucleus EOS based on the Virial expansion with multi-nucleus at low densities is switched to 
a single-nucleus EOS via  Hartree approximation at high densities, i.e., 
the multi-nucleus description is employed only in the low density regime and, as a result,
 some quantities are discontinuous or non-smooth at the transition between the two descriptions.  

We constructed 
in the previous papers
   \citep{furusawa11,furusawa13a}
an EOS based on the mass formula for nuclei under the influence of surrounding nucleons and electrons.
The mass formula extended from the 
 LDM   to describe  nuclear shell effects  as well as various in-medium effects, in particular, the formation of the pasta phases \citep{watanabe05,newton09,okamoto12}. 
The binding energies of $d$ (deuteron),  $t$ (triton), $h$ (${\rm{He}}^3$) and $\alpha$ (${\rm{He}}^4$) 
  are estimated with a quantum approach to light clusters \citep{typel10,roepke09}.  
Other light nuclei ($Z \leq 5$) are described with another
mass formula based on the LDM,
which is different from the one for heavy nuclei.
 The details of the model and comparisons with H. Shen's EOS  and HS EOS are given in \citet{furusawa11}.  In \cite{buyukcizmeci13},  three multi-nucleus EOS's (SMSM, HS and ours) were compared in detail. 
The extension of shell effects and
the implementation of the temperature dependence in bulk energies for heavy nuclei and the introduction of the quantum approach to $d$, $t$, $h$ and $\alpha$  were reported in \cite{furusawa13a}. 

The purpose of this study is to further improve the previous model, in which temperature dependences of binding  energies are lacked in part, and to discuss  their impact on EOS systematically.
Most significant change is the introduction of the washout of shell effects at high temperatures.
It is pointed out  by experimental and theoretical studies 
 that nuclear shell effects  disappear   completely at $T \sim$ 2.0~-~3.0~MeV
 \citep{brack74, bohr87, sandulescu97,nishimura14}. 
All supernova EOS's with multi-nucleus, however, have not considered this effect.
In SMSM EOS,  shell effects are ignored from the beginning  
and completely washed out  even at zero temperature.
Other multi-nucleus  EOS's including our previous model assume full shell effects at any temperature. 
The  shell effects are known to affect the nuclear abundance and electron capture rates of heavy nuclei  to a great extent \citep{buyukcizmeci13,raduta16}.
Other improvements 
 are the  introduction of the surface tension in its temperature dependence and
the reformulation of  model for light nuclei other than $d$, $t$, $h$ and $\alpha$. 
In the following, we report on these new ingredients and discuss the differences from the previous version.

This article is organized as follows.
In section 2, we describe  our new model with a focus on the  development from the previous EOS. 
Note that the basic formulation of the model free energy and its minimization are unchanged from the previous version.
The results  are shown in section 3, with an emphasis on the impact of the temperature dependence in binding energies. 
The paper is wrapped up with a summary and some discussions in section~4.
In appendix,  the data table of this EOS  is explained for use in supernova simulations.

\section{Formulation of the model}
To obtain the EOS with multi-nucleus, 
we construct a model free energy to be minimized with respect to the parameters included.
The supernova matter  consists of nucleons and nuclei together with electrons, photons and neutrinos.
Electrons and  photons are not treated in this paper, 
since their
inclusion 
as ideal Fermi and Bose gases, respectively, is trivial.
Note that the Coulomb energies between protons, both inside and outside nuclei, 
and electrons are contained in the EOS. We assume the electrons are uniformly distributed. Neutrinos are not always in thermal or chemical equilibrium with matter. We do not include them in  the free energies.  

The free energy of our model is constructed as a sum of the contributions  from free nucleons not bound in nuclei,
 light nuclei defined here as those nuclei with the proton number $Z\leq 5$,
and the rest of heavy nuclei  with the proton and neutron numbers, $Z\leq1000$ and $N\leq1000$.
We assume that the free nucleons outside nuclei interact with themselves only in the volume that is  not occupied by other nuclei; light nuclei are the quasi particles whose masses are modified by the surrounding free nucleons; heavy nuclei are also affected by the free nucleons and electrons, depending on the temperature and density. 
The free energy of free nucleons is calculated by the RMF theory with the excluded volume effect being taken into account. The model free energy of heavy nuclei is based on the liquid drop mass formula  taking  the following issues into account: the nuclear masses at low densities and temperatures should be equal to those of isolated nuclei  in vacuum;
one should take into account the effect that the nuclear bulk, shell, Coulomb and surface energies are affected by the free nucleons and electrons at high densities and temperatures;
furthermore the pasta phases near the saturation densities should be also accounted for to ensure a continuous transition to uniform matter.  The free energy of light nuclei is approximately calculated by quantum many body theory. 

In the following subsections, we explain the details of the free energy density, 
\begin{eqnarray}
\ f = f_{p,n}+\sum_{j}{n_j F_j } + \sum_i {n_i F_i}, \\
\ F_{j/i}=E^t_{j/i} + M_{j/i},
\end{eqnarray}
where $f_{p,n}$ is the free energy densities of free nucleons, $n_{j/i}$, $F_{j/i}$, $E^t_{j/i}$ and $M_{j/i}$ 
are the number density, free energy,  translational energy and rest mass of individual nucleus and 
 indices  $j$ and $i$ specify  light and
  heavy nuclei, respectively. 
The modifications from our previous EOS \citep{furusawa13a} are temperature dependences of shell and surface energies and the model of the light nuclei $(Z_j \leq 5)$ other than $d, t, h$ and $\alpha$ such as  $ ^7\rm{Li}$.
The other parts in free energies and the minimization of the total free energy densities are just the same as in the previous paper \citep{furusawa13a}. 

\subsection{minimization of free energy}
The thermodynamical quantities and abundances of nuclei as a function of $\rho_B$, $T$ and $Y_p$ are obtained by minimizing the model free energy with 
respect to the number densities of nuclei and nucleons under the constraints,
\begin{eqnarray}
 n_p+n_n+\sum_j{A_j n_j} +\sum_i{A_i n_i} & = & n_B=\rho_B/m_u, \nonumber \\
 n_p+\sum_j{Z_j n_j}+\sum_i{Z_i n_i} & = & n_e=Y_p n_B,
\label{eq:cons}
\end{eqnarray}
where $n_B$ and $n_e$ are number densities  of baryon and electrons, $A_{j/i}$ and $Z_{j/i}$ are the mass and proton numbers of nucleus $j/i$ and  $m_u$ is the atomic mass unit.
The minimization of our free energy density is not the same as that in the ordinary NSE,
in which  the number 
densities of all nuclei as well as nucleons are expressed by
two variable, i.e., the chemical potentials of nucleons $\mu_p$ and $\mu_n$ and 
one has only to solve  the constraints, Eq.~(\ref{eq:cons}) for the two variables, through Saha equations.
 In our case, on the other hand, the free energy density of nuclei depends on the local number densities of proton and neutron $n'_{p/n}$ to evaluate in-medium effects accurately 
and  $\mu_p$ and $\mu_n$ are expressed as follows:
\begin{equation}
\mu_{p/n} =\frac{\partial f}{ \partial n_{p/n}}= \mu'_{p/n}(n'_p,n'_n) + \sum_{i/j} n_{i/j} {\frac{\partial F_{i/j}(n'_p,n'_n)}{ \partial n_{p/n}}}, 
\label{eq:chem}
\end{equation}
where  $\mu'_{p/n}$ is the chemical potentials of nucleons in the vapor (see section~\ref{secnuc}) and the second term  is originated from the dependence of the free energies of nuclei on the nucleon densities
in the vapor (see sections~\ref{sechn}  and \ref{secln}) and hence is summed over all nuclear species.
Thus the number densities of nuclei are not determined by $\mu_p$ and $\mu_n$ alone but they also depend on $n'_p$ and $n'_n$.
We hence solve  the equations relating $\mu_{p/n}$ and $n'_{p/n}$, Eq.~(\ref{eq:chem}), 
 as well as the two constraint equations, Eq.~(\ref{eq:cons}),
to determine the four variables: $\mu_p$, $\mu_n$, $n'_p$ and $n'_n$.

\subsection {free energy for nucleons outside nuclei  \label{secnuc}}
The free energy density of free nucleons is calculated by the RMF theory with the TM1 parameter set,  which is determined
so that it should reproduce the properties of heavy nuclei in the wide mass range including neutron rich nuclei, and is the same as that adopted in \citet{shen11}.
We take into account the excluded-volume effect:
 free nucleons can not move in the volume occupied by other nuclei, $V_N$.
Then the local number densities of free protons and neutrons are defined as $n'_{p/n}=(N_{p/n})/(V-V_N)$  with the total volume, $V$, and the numbers of free protons, $N_p$, and  free neutrons, $N_n$.
Then the free energy densities  of free nucleons are defined as $f_{p,n} =(V-V_N)/V \times  f^{RMF}(n'_p,n'_n,T)$,
where $f^{RMF}(n'_p,n'_n,T)$ is the free energy density in the unoccupied volume for nucleons, $V-V_N$,  obtained from the RMF theory at $n'_p$, $n'_n$  and temperature $T$.

\subsection {masses of heavy nuclei ($Z\geq 6$) \label{sechn}}
The nuclear mass is assumed to be the sum of  bulk, Coulomb, surface  and shell energies: $M_i=E_i^B+E_i^C+E_i^{Su} +E_i^{Sh}$.
The formulation of bulk and Coulomb energies is just identical  to the previous one.
In the new model, we take into account the temperature dependences of shell and surface energies. 

We define the saturation densities of nuclei $n_{si}(T)$ 
as the baryon number density, at which the free energy per baryon
 $F^{RMF}(T,n_B, Y_p)$ given by the RMF with $Y_p=Z_i/A_i$ takes its minimum value. 
Thus $n_{si} (T)$ depends on the temperature $T$ and the proton fraction in each nucleus $Z_i/A_i$.
At temperatures larger than a critical temperature $T_{ci}$, the free energy, $F^{RMF}(T,n_B, Z_i / A_i)$, has no minimum 
because of the entropy contribution. 
Then the saturation density $n_{si} (T)$ above $T_{ci}$  is  assumed to be equal to the saturation density at the critical temperature $n_{si}(T_{ci})$. 
When the saturation density $n_{si}$ so obtained  is lower than the baryon number density of the whole system $n_B$, we reset the saturation density as the baryon number density $n_{si}=n_B$.
This prescription approximately represents compressions of nuclei near the saturation densities. These treatments of the saturation density are important in obtaining reasonable bulk energies at high temperatures and densities \citep{furusawa11,furusawa13a}. 
The bulk energies are evaluated from the free energy per baryon of the uniform nuclear matter  at the saturation density $n_{si}$ for a given temperature $T$ and proton fraction inside the nucleus $Z_i/A_i$ as
$E_{i}^{B} =A_i \{M_B +  F^{RMF}(T,n_{si},Z_i/A_i) \}$, 
where the baryon mass  $M_B$ is set to be 938 MeV.

The Wigner-Seitz cell (W-S cell) for each species of  nuclei is set to satisfy charge neutrality 
 with the volume, $V_i$.
The cell also contains free nucleons as a vapor outside the nucleus as well as  electrons distributed  uniformly in the entire cell.
The charge neutrality in the cell gives the cell volume $V_i = (Z_i - n'_p  V_i^N)/(n_e-n'_p)$ 
where $V_i^N$ is the volume  of the nucleus in the cell and can be calculated as $V_i^N = A_i / n_{si}$. 
The vapor volume and nucleus volume fraction in the cell are given by $ V_i^B = V_i-V_i^N  $ and $u_i =  V_i^N / V_i$, respectively.
In this EOS, we assume that each nucleus enters the nuclear pasta phase individually when 
the volume fraction reaches $u_i=0.3$ and that the bubble shape is realized when it exceeds $0.7$. 
The bubbles are explicitly treated as  nuclei of spherical shell shapes with the vapor nucleons filling the inside.
This phase is important to ensure continuous transitions to uniform matter as noted in \citet{furusawa11}.
The intermediate states ($ 0.3<u_i<0.7 $) are smoothly interpolated from the normal and bubble states. 
The evaluation of the Coulomb energy in the W-S cell is given by the integration of Coulomb forces in the cell:
\begin{eqnarray}
E_i^C=\left\{ \begin{array}{ll}
\displaystyle{\frac{3}{5}\left(\frac{3}{4 \pi}\right)^{-1/3}  e^2 n_{si}^2 \left(\frac{Z_i - n'_p V_i^N}{A_i}\right)^2 {V_i^N}^{5/3} D(u_i)}    & (u_i\leq 0.3), \\
\displaystyle{\frac{3}{5}\left(\frac{3}{4 \pi}\right)^{-1/3}   e^2 n_{si}^2 \left(\frac{Z_i - n'_p V_i^N}{A_i}\right)^2 {V_i^B}^{5/3} D(1-u_i)}   & (u_i\geq 0.7),
\end{array} \right.
\label{eqclen}
\end{eqnarray} 
with $D(u_i)=1-\frac{3}{2}u_i^{1/3}+\frac{1}{2}u_i$, where $e$ is the elementary charge. 

The surface energies are evaluated as 
\begin{eqnarray} \label{eq:surf}
E_i^{Su}&=&\left\{ \begin{array}{ll}
4 \pi {r^2_{Ni}} \, \sigma_i \left(1-\displaystyle{\frac{n'_p+n'_n}{n_{si}}} \right)^2 \left( \displaystyle{\frac{T^2_c-T^2}{T^2_c+T^2}} \right)^{5/4}   & (u_i\leq 0.3), \\
4 \pi {r^2_{Bi}} \, \sigma_i  \left(1-\displaystyle{\frac{n'_p+n'_n}{n_{si}}} \right)^2 \left( \displaystyle{\frac{T^2_c-T^2}{T^2_c+T^2}} \right)^{5/4}  & (u_i\geq 0.7), 
\end{array} \right.\\
&&\sigma_{i}=\sigma_0  - \frac {A_i^{2/3} } {4 \pi r_{Ni}^2} [S_s(1- 2 Z_i/A_i)^2 ],
\end{eqnarray} 
where  $r_{Ni} = ( 3/4 \pi V^N_i)^{1/3}$ and $r_{Bi} = ( 3/4 \pi V^B_i)^{1/3}$ are the radii of nucleus and bubble and 
$\sigma_0$ denotes the surface tension for symmetric nuclei. 
The surface tension  $\sigma_i$  includes the surface symmetry energy,
 i.e., neutron-rich nuclei have lower surface tensions than symmetric nuclei.
 The values of the constants, $\sigma_0=1.15$ MeV/fm$^3$ and 
$S_s =$ 45.8 MeV, are adopted from the paper by \cite{lattimer91}. 
The factor in Eq.~(\ref{eq:surf}), $ \left(1-(n'_p+n'_n)/n_{si} \right)^2 $, is
added by hand to take into account the effect that the surface energy should be 
reduced as the density contrast decreases between the nucleus and the nucleon vapor.
The last factor in  Eq.~(\ref{eq:surf}), $((T^2_c-T^2)/(T^2_c+T^2))^{5/4} $,  accounts
for the temperature dependence in the surface tension  and is  one of the improvements in the new model. This factor reproduces approximately  the  liquid-gas  phase  transition   
at   the critical temperature $T_c=18$ MeV, where the nuclei are dissolved to free nucleons \citep{ravenhall83,bondorf95}.
This factor is also adopted in the SMSM EOS \citep{botvina04, botvina10,buyukcizmeci14}.
Note that the values of $\sigma_i$ and  $T_c$ are not consistent with 
 the RMF theory with the TM1 parameter set 
and the dependence of  $T_c$ on the proton fraction in each nucleus $Z_i/A_i$ is not taken into account for simplicity. 
We use cubic polynomials of $u_i$ for the interpolation between the droplet and bubble phases. 
The four coefficients of the polynomials are determined by the condition that the Coulomb and surface energies are continuous and smooth as a function of $u_i$ at $u_i=0.3$ and $u_i=0.7$.

We include the shell effects separately in the mass formula of nuclei by the use of
both experimental and theoretical mass data \citep{audi12, koura05} to better reproduce the nuclear abundances in the low density regime.
The shell energies are obtained from the mass data  by subtracting our liquid drop mass formula, which does not include the shell effects, $(M_i^{LDM}=E_i^B+E_i^C+E_i^{Su})$ in the vacuum limit with  $T, n'_{p/n}, n_e =0$ as $E_{i0}^{Sh} =M_i^{data} -[ M_i^{LDM} ]_{vacuum}$.
Note that the shell energy in our mass formula actually includes pair
energies.
We take the in-medium effect into account phenomenologically as follows:
\begin{eqnarray}\label{eq:sh}
 E_{i}^{Sh}(\rho,T)&=&\left\{ \begin{array}{ll}
E_{i0}^{Sh} \displaystyle{\frac{\tau_i}{{\rm sinh}\tau_i}} & (\rho \leq 10^{12} \rm{g/cm^3}), \\
\\
E_{i0}^{Sh} \displaystyle{\left( \frac{\tau_i}{{\rm sinh}\tau_i}\right) \left( \frac{\rho_0- \rho}{\rho_0 - 10^{12}  \rm{g/cm^3}} \right) }& (\rho >  10^{12} \rm{g/cm^3}), 
\end{array} \right.
\end{eqnarray} 
where $\rho_0$ is taken to be $m_u$ times  the saturation density of symmetric nuclei $n_{si}(T, Z_i/A_i=0.5)$ at temperature $T$.
The  factor $\tau_i/{\rm sinh}\tau_i$ expresses the washout of shell effects approximately, which is derived   by the analytical study for the  single particle motion of nucleon outside the closed shell \citep{bohr87}. 
The normalized factor $\tau_i$ is defined as $\tau_i=2 \pi^2 T/\epsilon_i^{sh}$
with the energy spacing of  the shells, $\epsilon_i^{sh} = 41 A_i^{-1/3}$~MeV.
Note that  $E_0^{Sh}$ and $\epsilon_i^{sh}$ are not based 
on exact calculations of  nuclear structures 
but the simple assumptions 
and  in this sense, this formulation of the washout  is an approximation.  
The fully self-consistent calculation with the nuclear composition and structures of all nuclei  including in-medium effects is much beyond the scope of this study.
It is noted that  this approximate formulation  can reproduce qualitatively the feature of washout that the shell effects disappear around $T\sim$~2.0-3.0 MeV as shown later in Sec~\ref{results}.
The linear interpolation, $(\rho_0- \rho_B)/(\rho_0 - 10^{12} \rm{g/cm^3})$, accounts for the decay of shell effects at high densities because of the existence of electrons, free nucleons and other nuclei. 
This disappearance of shell energies  is essential for our EOS model to reproduce continuous transition to the uniform nuclear matter at saturation densities \citep{furusawa11}. 
The choice  of the critical density $10^{12} \rm{g/cm^3}$  is rather arbitrary, 
since the dependence of the shell energies on the density  of ambient matter has not been thoroughly investigated yet.
  It is noted, however, that the
structure of nuclei is known to be affected by ambient matter at
these densities \citep{aymard14}.

\subsection {masses of light nuclei ($Z\leq 5$) \label{secln}}
We describe 
light nuclei
as  quasi-particles outside heavy nuclei. 
Their masses   are assumed to  be given by the following expression:
\begin{equation}
  M_{j}=M_j^{data} + \Delta E_{j}^{Pa}  + \Delta E_j^{SE} +\Delta E_j^C \ \ \ \ (j=d,t,h \  \&  \ \alpha),  \label{eqquasi}
\end{equation}
where $\Delta E_j^{Pa}$ is the Pauli energy shift by other baryons, 
$\Delta E_j^{SE}$ is the self-energy shift of the nucleons composing the light nuclei and  $\Delta E_j^C$ is the Coulomb energy shift.

For the Pauli energy shifts of  $d, t, h$ and $\alpha$, we employ the empirical formulae provided by \citet{typel10},
 which are quadratic functions fitted to the result of quantum statistical calculations \citep{roepke09}.
The local proton and neutron number densities that include light nuclei are defined as
\begin{eqnarray}
n_{pl}=n'_p + \eta^{-1} \sum_{j}  {Z_j n_j} , \\ 
n_{nl}=n'_n + \eta^{-1} \sum_{j}  {N_j n_j}  \ . 
\end{eqnarray}
where $\eta$ stands for the volume fraction $(V-V_N)/V$. 
Then  the Pauli energy shift $\Delta E_j^{Pa}$ is given by the following expression \citep{typel10}:
\begin{eqnarray}
\label{DEP}
& & \Delta E_{j}^{Pa}(n_{pl},n_{nl},T)  =  - \tilde{n}_{j} 
 \left[1 + \frac{\tilde{n}_{j}}{2\tilde{n}_{j}^{0}(T)} \right]
 \delta B_{j} (T), \\
 \delta B_{j}(T) &=&\left\{ \begin{array}{ll}
  a_{j,1} / T^{3/2}  \left[ 1/\sqrt{y_{j}}  -  \sqrt{\pi}a_{j,3} \exp  \left( a_{j,3}^{2} y_{j} \right) 
 {\rm erfc} \left(a_{j,3} \sqrt{y_{j}} \right) \right]   &  {\rm{for}} \   j  =  d, \\
  a_{j,1} /\left( T  y_j \right)^{3/2}  \ \   &  {\rm{for}} \  j  =  t,h,\alpha, 
\end{array} \right.
\end{eqnarray}
where  $\tilde{n}_{j} = 2(Z_{j} \ n_{pl} +N_{j} \ n_{nl}) /A_{j}$ and $y_{j} = 1+a_{j,2}/T$.
The density scale for the dissolution of each light nucleus is given by  $\tilde{n}_{j}^{0}(T) = B_{j}^{0}/\delta B_{j}(T)$
with the binding energy in vacuum, $B_{j}^{0}=Z_j M_p+N_j M_n -M_j^{data}$. 

The self-energy shifts of light nuclei are the sum of
 the self-energy shifts of individual nucleons composing the light nuclei  $\Delta E^{SE}_{n/p}=\Sigma^0_{n/p}(T,n'_p,n'_n)-\Sigma_{n/p}(T,n'_p,n'_n)$ with  $\Sigma^0$ and $\Sigma$ being the vector and scalar potentials of nucleon and 
the contribution from their effective masses $\Delta E_{j}^{\rm eff.mass}$:
\begin{equation}
\label{SE}
\Delta E_{j}^{SE}(n'_{p},n'_{n},T)= (A_j-Z_j) \Delta E_{n}^{SE}+ Z_j \Delta E_{p}^{SE} +\Delta E_{j}^{\rm eff.mass}\ . 
\end{equation}
The latter is 
 given as
$\Delta E_{j}^{\rm eff.mass} = \left(1-m^{\ast}/m \right)s_{j}$
with $m^{\ast}=m_B-\Sigma_{n/p}(T,n'_p,n'_n)$.
The potentials $\Sigma^0$ and $\Sigma$ are calculated from the RMF employed for free nucleons as noted in \ref{secnuc}.
More detailed explanations of the Pauli- and self-  energy shifts of  $d, t, h$ and $\alpha$ are provided in \citet{typel10,roepke09,furusawa13a}.
The parameters $a_{j/1}$, $a_{j/2}$, $a_{j/3}$, and $s_j$ are given in Table~1 in \cite{furusawa13a}. 

There is little information about the  in-medium effects for the light nuclei $(Z_j \leq 5)$ other than  $d$, $t$, $h$ and $\alpha$.
In the previous model,  their binding energies are evaluated with the LDM that
is different  from the one for heavy nuclei.
Their bulk energies are linearly interpolated between the two different estimations  at $\rho_B=10^{12}$~g/cm$^3$ and at the nuclear saturation density:
at the lower end, 
they are evaluated  from the experimental or theoretical mass data in the vacuum limit;
at the other end they are designed to agree with 
  the bulk energies 
for heavy nuclei in the same LDM.
The  light nuclei are assumed to form nuclear pastas together with heavy nuclei ($Z \geq 6$) near the saturation densities. 
It is noted, however, that there is no justification for this interpolation. 
For simplicity these light nuclei are assumed to be quasi particles 
 in the same way as $d, t, h$ and $\alpha$ at all sub-nuclear densities 
and their mass are  evaluated with 
Eq.~(\ref{eqquasi}) in the new model. 
The Pauli-energy shifts for  them  are calculated
with Eq.~(\ref{DEP}) in the same way as that  for  $\alpha$. The self-energy shifts for them are set to be zero, on the other hand. 
This reformulation has 
only minor influence on the nuclear composition and thermodynamical quantities, since the light nuclei other than  $d$, $t$, $h$ and $\alpha$
 do not become  abundant as shown later. 
This is because 
they have  smaller binding energies 
 than $\alpha$
and are less populated than
 $d$, $t$, $h$ and $\alpha$ 
 at high temperatures, which give higher entropies per baryon.

The Coulomb energy shifts are calculated as
\begin{equation}
 \Delta E_j^{C} = E_j^C(n'_p,u_j)-E_j^C(0,0), 
\end{equation}
where $E_j^C$  is given as
\begin{equation} 
 E_j^C (n'_p,u_j) = \frac{3}{5}\left(\frac{3}{4 \pi}\right)^{-1/3}  e^2 n_{sj}^2 \left(\frac{Z_j - n'_p V_j^N}{A_j}\right)^2 {V_j^N}^{5/3} D(u_j). 
\end{equation} 
Although the expression of the Coulomb energy is identical to that for heavy nuclei,
 the shifts are negligible compared with other energy shifts.
We do not take into account the nuclear pasta phases and surface- and shell-energies both for  $d$, $t$, $h$ and $\alpha$ and for the rest of light nuclei.

\subsection{Translational energies of nuclei} \label{sectr}
The translational energy of nucleus $i$ in our model free energy is based on that for the ideal Boltzmann gas and given by 
\begin{equation}%
\label{eq:tra}
 F_{i/j}^{t}=k_B T \left\{ \log \left(\frac{n_{i/j}}{g_{i/j}(T) n_{Qi/j}}\right)- 1 \right\} \left(1-\frac{n_B}{n_s}\right), 
\end{equation}
where $k_B$ is the Boltzmann constant and $n_{Qi/j} = \left(M_{i/j} k_B T/2\pi \hbar ^2  \right)^{3/2}$.
We modify the internal degree of freedom $g_{i/j}(T)$ for heavy nuclei 
in this paper so that it should  approach unity  in line with the  washout of the shell effects:
 $g_i(T)= (g_i^0 -1)\displaystyle{\frac{\tau_i}{{\rm sinh}\tau_i}}  +1$. 
In the previous model, we just used  the spin degree of freedom of the ground state as  $g_i(T)= g_i^0$ at any temperature.
  Note that the contribution of the excited states to free
energy is included in the temperature dependence of the
bulk energy. 
As for light nuclei, $g_j^0$ is always adopted.
The last factor on the right hand side of Eq.~(\ref{eq:tra}) takes account of the excluded-volume effect: each nucleus can move in 
the space that is not occupied by other nuclei and free nucleons in the same way as in \cite{lattimer91}.
The factor reduces the translational energy at high densities and
is important to ensure the continuous transition to uniform nuclear matter.
We always employ the nuclear saturation density of symmetric matter $n_s=\left[ n_{si}(Z_i/A_i,T) \right] _{Z_i/A_i=0.5}$ in Eq.~(\ref{eq:tra}) for numerical convenience.

\subsection{Thermodynamical quantities \label{secth}}%
By minimization, we obtain the free energy density together with the abundances of all nuclei 
and free nucleons as a function of $\rho_B$, $T$ and $Y_p$. 
Other physical quantities are derived by partial differentiations of the optimized free energy density.
All the terms for the 
 in-medium effects
are properly taken into account in this process
 to ensure the thermodynamical consistency as described in \citet{furusawa11} in detail.
The baryonic pressure, for example, is obtained by the differentiation with respect to the baryonic density
as follows:  
\begin{eqnarray}
 p_B &=&n_B \left[\partial{f}/\partial {n_B}\right]_{T,Ye}-f,  \nonumber \\
   &=&   p_{p,n}^{RMF}+ \sum_{i/j} ( p_{i/j}^{th}  + p_{i/j}^{ex}  + p_{i/j}^{mass}), \label{eq:pr} \\ 
   p_{i}^{mass}&= & (p_{i}^{shell} + p_{i}^{Coul} + p_{i}^{Surf}) \ \ \ \  \ \ \  (\rm{heavy  \ nuclei} \ \it{Z_i} \geq \rm{6}), \\ 
   p_{j}^{mass}&= & (p_j^{Pauli} + p_j^{SE} + p_j^{Coul})  \ \ \ \ \  \  \ \ \ \ (\rm{light  \ nuclei} \ \it{Z_j} \leq \rm{5}),    \label{eq:prl} 
\end{eqnarray}
where $ p_{p,n}^{RMF}$ is the contribution of the nucleons in the vapor; 
both $p_{i/j}^{th}=n_{i/j} k_B T (1-n_B/n_s)$  and $p_{i/j}^{ex}=n_{i/j} k_B T (n_B/n_s)({\rm log}(n_{i/j}/n_{Qi/j})-1)$ come from the translational energy of nuclei in the free energy;
$p_{i}^{shell}$, $p_{i}^{Coul}$ and $ p_{i}^{Surf}$ 
originate from the shell, Coulomb and surface energies of heavy nuclei in the free energy, respectively;
 $p_j^{Pauli}$, $p_j^{SE}$ (for  $d$, $t$, $h$ and $\alpha$)  and  $p_j^{Coul}$ are derived from the Pauli-, self- and Coulomb-energy shifts of the light nuclei.

The entropy per baryon is calculated from the following expression:
\begin{eqnarray}
 s& =& -\frac{\left[\partial{f}/\partial {T}\right]_{\rho_B,Ye}}{n_B},  \label{eq:ent} \\
    &=&\eta s^{RMF}_{p,n} +  \sum_{i/j} \frac{n_{i/j} k_B}{n_B} \Biggl[ \left\{  \frac{5}{2}- \log \left(\frac{n_{i/j}}{g_{i/j} n_{Qi/j}} \right)  + \frac{\partial{g_i(T)}}{\partial {T}} /g_i(T)
   \right\} (1-n_B/n_s) 
-   \frac{\partial{M_{i/j}}}{\partial {T}}  \Biggl]  . \nonumber
\end{eqnarray}
In the new model,  the introduction of  temperature-dependence of some in-medium effects
yields  the term of  $\partial{g_i(T)}/\partial{T} /g_i(T)$ for heavy nuclei  and contributes to the partial derivative of the masses,
which are obtained as follows: 
\begin{eqnarray}
\label{eq:ent2}
 \frac{\partial{M_i}}{\partial {T}}  = &  - A_i s_i^{RMF} (T,n_{si}, Z_i / A_i)  +\frac{\displaystyle \partial{E_i^{Su}}}{\displaystyle \partial {T}} +\frac{\displaystyle \partial{ E_i^{Sh}}}{\displaystyle \partial {T}} 
 &\ \  (\rm{heavy  \ nuclei}),  \\
 \frac{\partial{M_j}}{\partial {T}} =  &\frac{\displaystyle \partial{\Delta E_j^{SE}}}{\displaystyle \partial {T}} + 
\frac{ \displaystyle \partial{\Delta E_j^{Pa}}}{ \displaystyle \partial {T}}
 \ \ \ \   (\rm{light  \ nuclei} \ \it{Z_j} \leq \rm{5}), & 
\end{eqnarray}
where the entropy per baryon $s_{i}^{RMF}$ is obtained by the RMF.
The contribution of this term is normally negligible except near the nuclear saturation density.

\section{Result} \label{results}
In this paper, we construct the EOS that improves
the previous one \citep{furusawa13a}.
The changes  are the additions of the temperature dependences in the surface and shell energies  of heavy nuclei and the reformulation of the mass evaluation  for the light nuclei other than $d$, $t$, $h$ and $\alpha$.
We compare the results of the different models  listed  in Table~\ref{tab1_model},
focusing 
 on  the impact of  the improvements. 
Model~0f is the same EOS as the previous EOS \citep{furusawa13a}, in which the binding energies of light nuclei $Z \leq 5$  other than $d$, $t$, $h$ and $\alpha$ are calculated  via the LDM.
In other models, they are evaluated with the quasi-particle method  as noted in Sec.~\ref{secln}.
The temperature dependence in the surface energy given in Eq.~(\ref{eq:surf}) with 
the last factor $((T^2_c-T^2)/(T^2_c+T^2))^{5/4} $ is employed in Models  2f, 2w  and 2n, 
whereas
the factor is dropped in Models 0f and  1f.
Models 0f,  1f and 2f ignore the washout of the shell effects, 
dropping the factor $\tau_i/{\rm sinh}\tau_i$ in Eq.~(\ref{eq:sh}). 
They may be regarded as a surrogate for HS EOS \citep{hempel10}  and the   ordinary NSE EOS \citep{timmes99}, 
 which also neglect the washout effect.
Model 2w is the new EOS, in which we take this  effect. 
We also prepare Model 2n for comparison, which do not consider the
shell effects at all  ($E_{i0}^{Sh}=0$).
This model is similar to the SMSM EOS, in which nuclei are described  by the LDM description without shell effects.
Note that the density dependence in the shell effects between $\rho_B=10^{12} \rm{g/cm^3}$ and $\rho_0$ in  Eq.~(\ref{eq:sh})  is included in all the model  although it is effective only near the saturation densities.  

First, we will elucidate
the difference  between Models 0f and 1f, paying attention to 
 the mass fraction of light nuclei. 
Then we discuss the impact of the temperature dependence of in-medium effects on the abundance of the heavy nuclei, based on the results of  Models 1f, 2f, 2w and 2n.
Finally, the thermodynamical quantities are compared for all models.

\begin{figure}
\includegraphics[width=8cm]{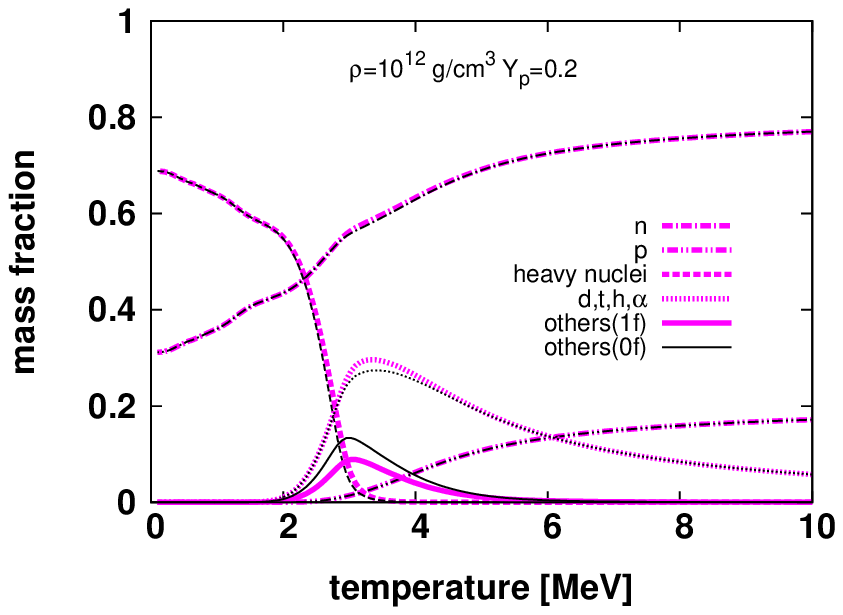}
\includegraphics[width=8cm]{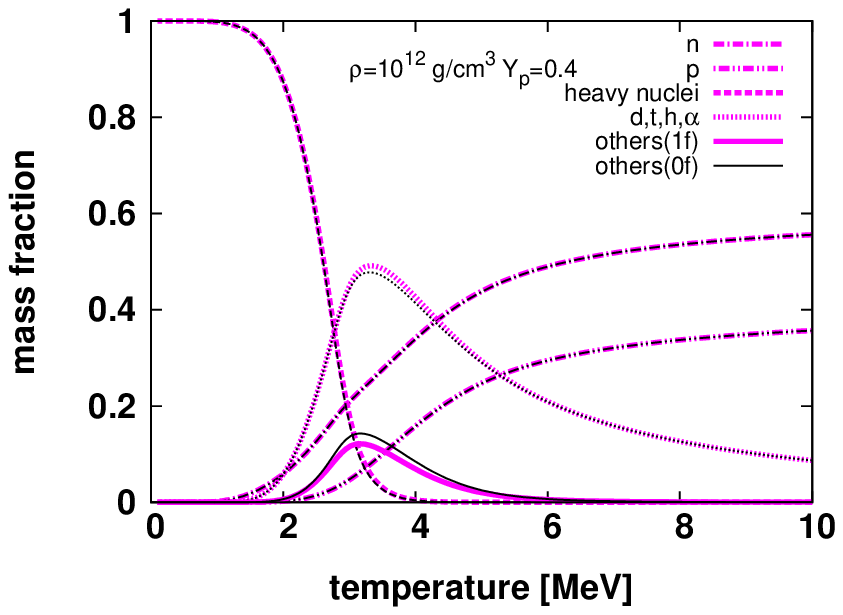}
\includegraphics[width=8cm]{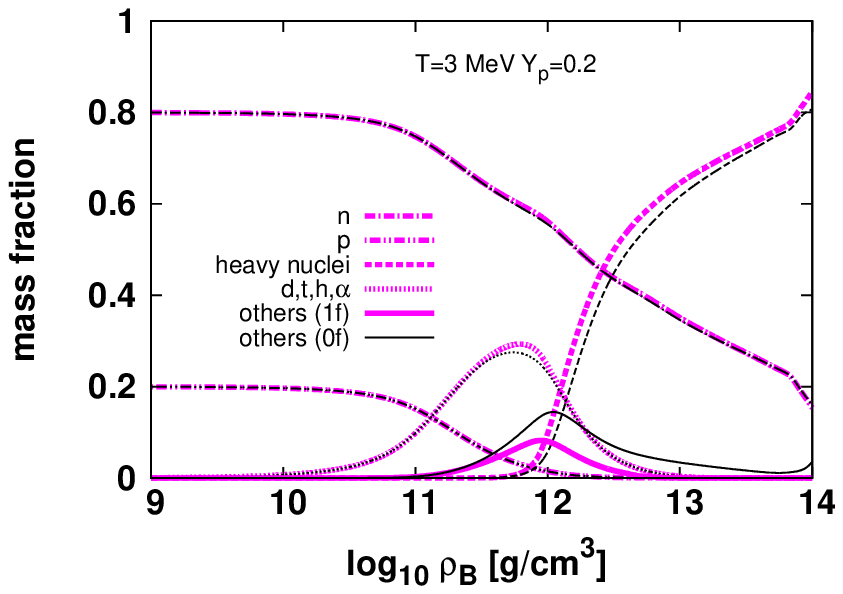}
\includegraphics[width=8cm]{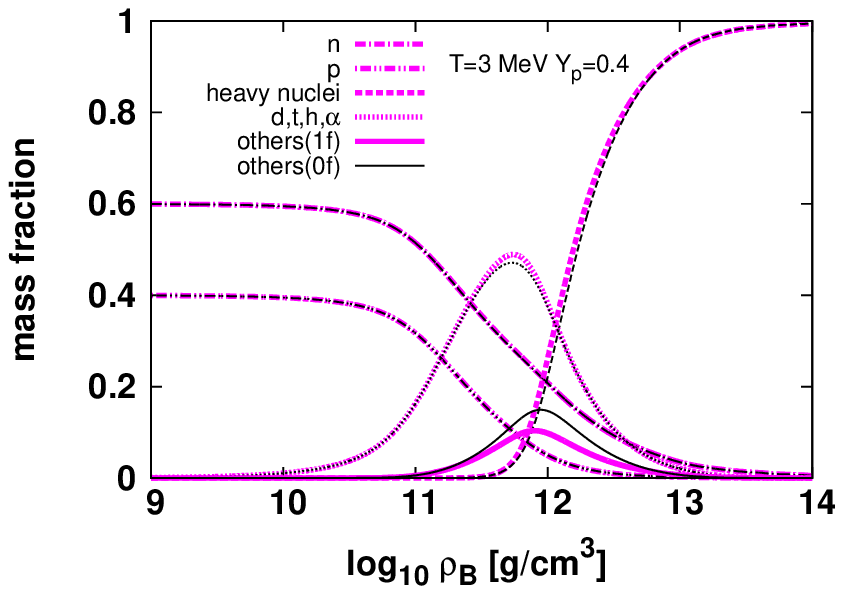}
\ \\
\ \\
\ \\
\caption{Mass fractions of  neutrons (dash dotted line), protons (dashed double-dotted lines), heavy nuclei with $Z \geq 6$(dashed lines),  $d$,  $t$, $h$, $\alpha$ (dotted lines), the rest of light nuclei  with $Z \leq 6$ (solid lines)  for Models~1f (magenta lines) and 0f  (black  lines), respectively, 
 at  $Y_p=$ 0.2 (left top panel) and 0.4 (right top panel) and  $\rho_B=10^{12} $ g/cm$^3$ as a function of temperature  as well as at  $T=$ 3~MeV and   $Y_p=$ 0.2 (left bottom panel) and  0.4 (right bottom panel)  as a function of density.
}
\label{fig_masslight}
\end{figure}

\begin{figure}
\includegraphics[width=8cm]{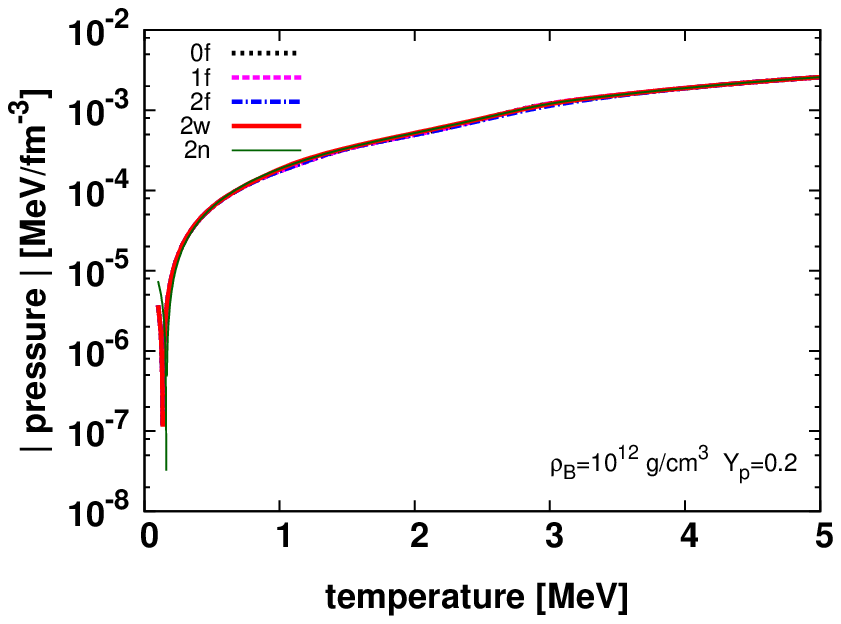}
\includegraphics[width=8cm]{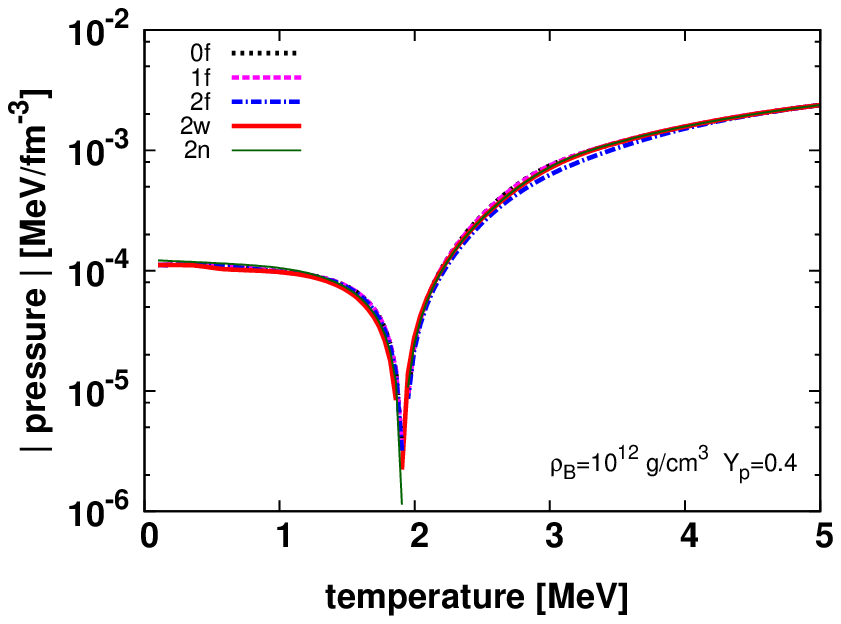}
\includegraphics[width=8cm]{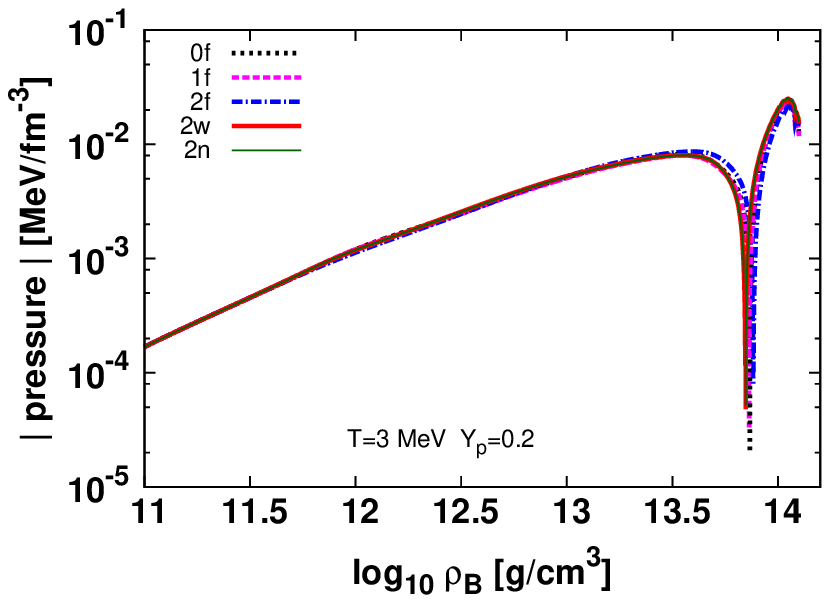}
\includegraphics[width=8cm]{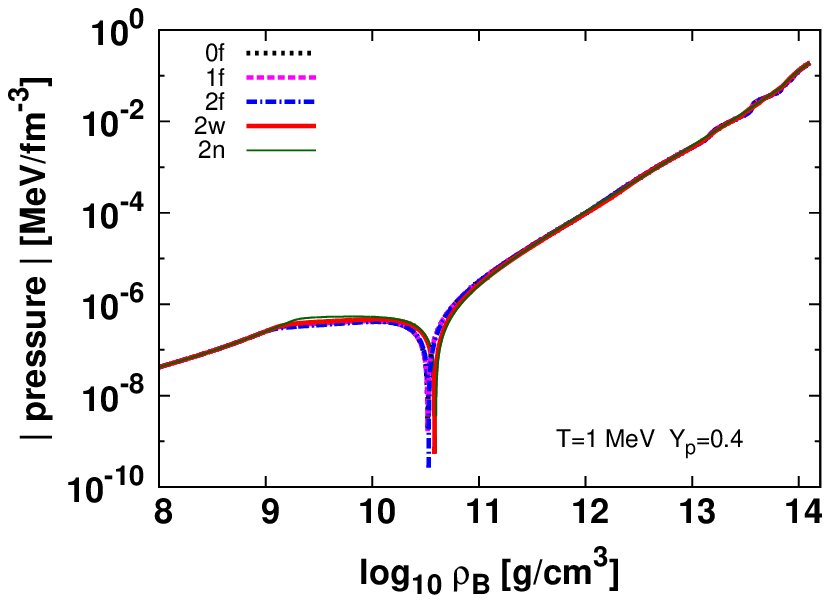}
\ \\
\ \\
\ \\
\caption{Absolute value of baryonic pressure for Models~0f (black dotted lines), 1f (magenta dashed lines), 2f  (blue dash dotted  lines), 2w  (red solid thick lines), 2n  (green solid thin lines)
 at  $Y_p=$ 0.2 (left top panel) and 0.4 (right top panel) and  $\rho_B=10^{12} $ g/cm$^3$ as a function of temperature  as well as at  $T=$ 3~MeV and     $Y_p=$ 0.2 (left bottom panel) and   $T=$ 1~MeV  and  $Y_p=$ 0.4 (right bottom panel)  as a function of density.
}
\label{fig_pre}
\end{figure}

\begin{figure}
\includegraphics[width=8cm]{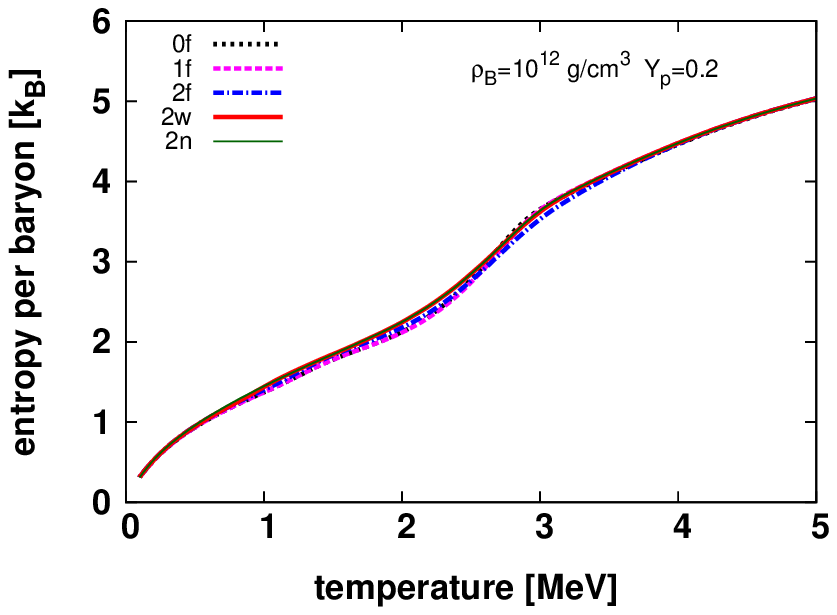}
\includegraphics[width=8cm]{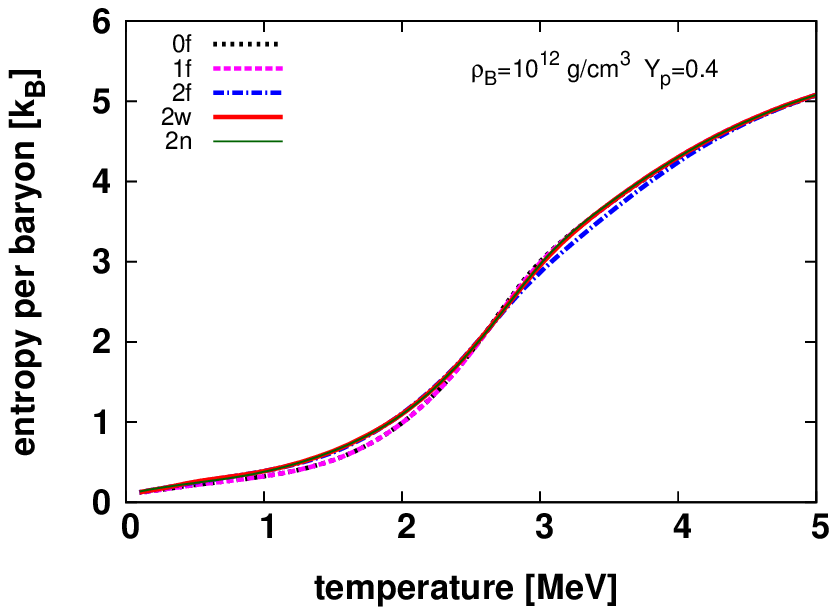}
\includegraphics[width=8cm]{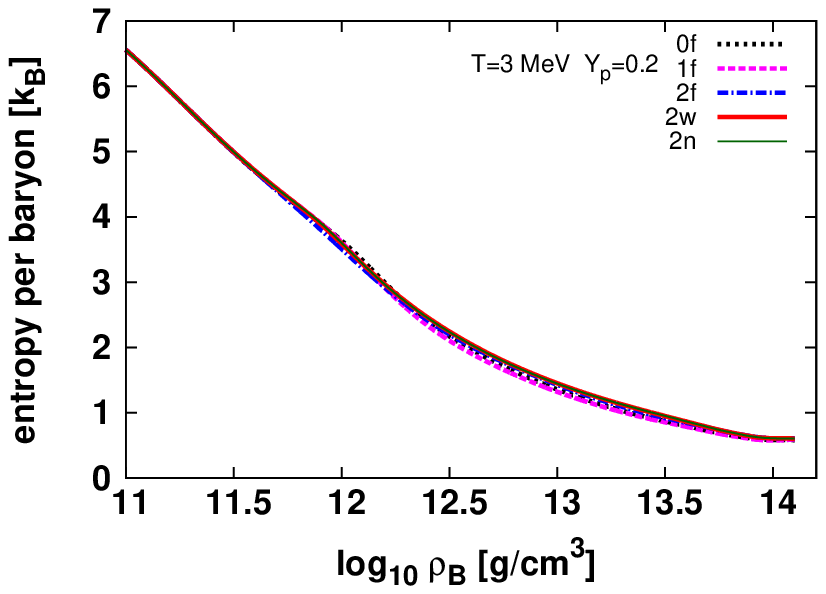}
\includegraphics[width=8cm]{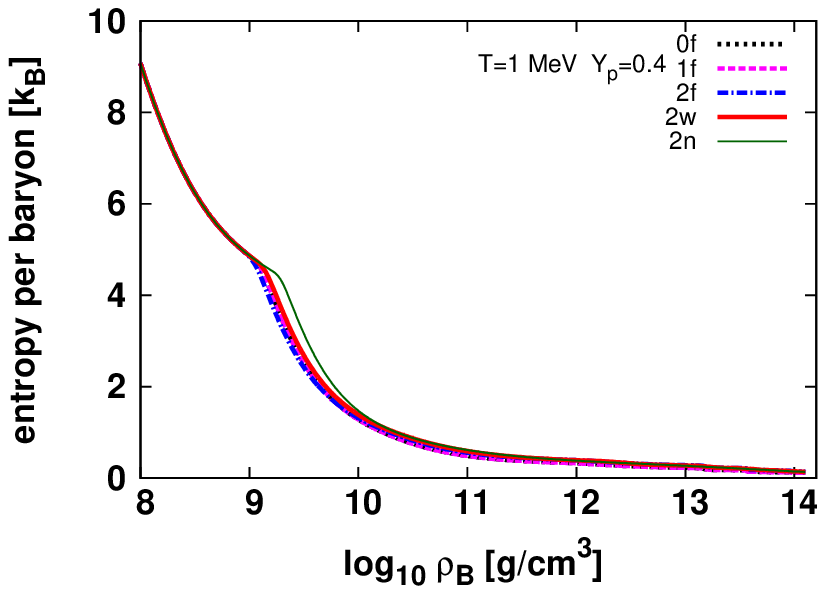}
\ \\
\ \\
\ \\
\caption{Entropy per baryon for Models~0f (black dotted lines), ~1f (magenta dashed lines), 2f  (blue dash dotted  lines), 2w  (red solid thick lines), 2n  (green solid thin lines)
 at  $Y_p=$ 0.2 (left top panel) and 0.4 (right top panel) and  $\rho_B=10^{12} $ g/cm$^3$ as a function of temperature  as well as at  $T=$ 3~MeV and     $Y_p=$ 0.2 (left bottom panel) and   $T=$ 1~MeV  and  $Y_p=$ 0.4 (right bottom panel)  as a function of density.
}
\label{fig_ent}
\end{figure}

Figure~\ref{fig_masslight} displays the mass fractions of proton, neutron, the light nuclei of $d$, $t$, $h$ and $\alpha$, the rest of light nuclei ($Z \leq 5$) as well as  heavy nuclei ($Z \geq 6$) for Models 0f and 1f 
 at $\rho_B=$~10$^{12}$~g/cm$^3$ and  $Y_p=$0.2 and 0.4  as a function of  temperature 
and those at  $T$=~3 MeV and  $Y_p=$0.2 and 0.4 as a function of density.
It is found that the mass fraction of the light nuclei other than 
$d$, $t$, $h$ and $\alpha$ decreases in Model 1f compared with Model 0f  due to the Pauli shifts.
The reduction does not affect the mass fractions of free nucleons and  the other nuclei take their place. 
The difference is apparently small, since these light nuclei are never dominant as already explained. 
The light nuclei in the pasta phase are also replaced by the heavy-nuclei pasta around  $\rho\sim 10^{14}$~g/cm$^3$   for  $T=3$~MeV and $Y_p=0.2$. 
These changes have essentially
no influence on thermodynamical quantities as shown later in Figs.~\ref{fig_pre} and \ref{fig_ent}. 

\begin{figure}
\includegraphics[width=8cm]{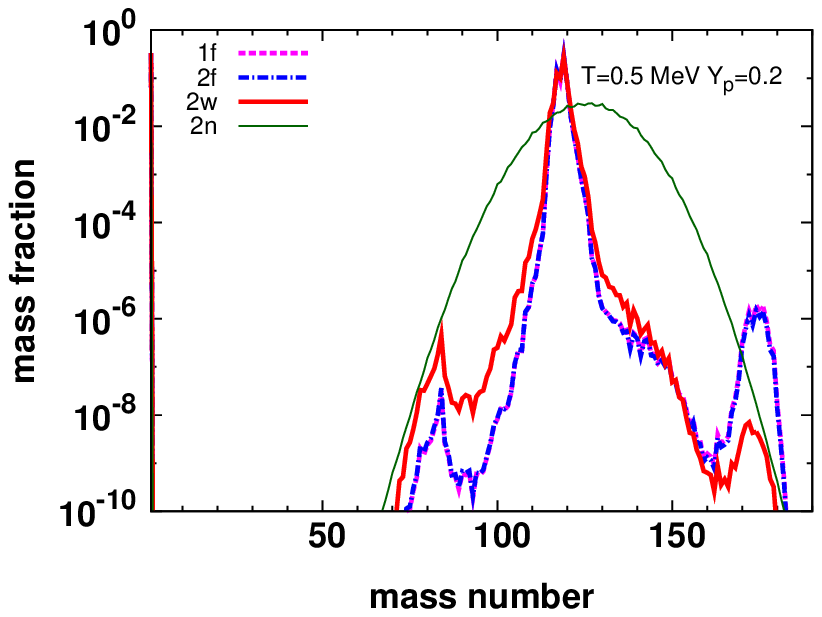}
\includegraphics[width=8cm]{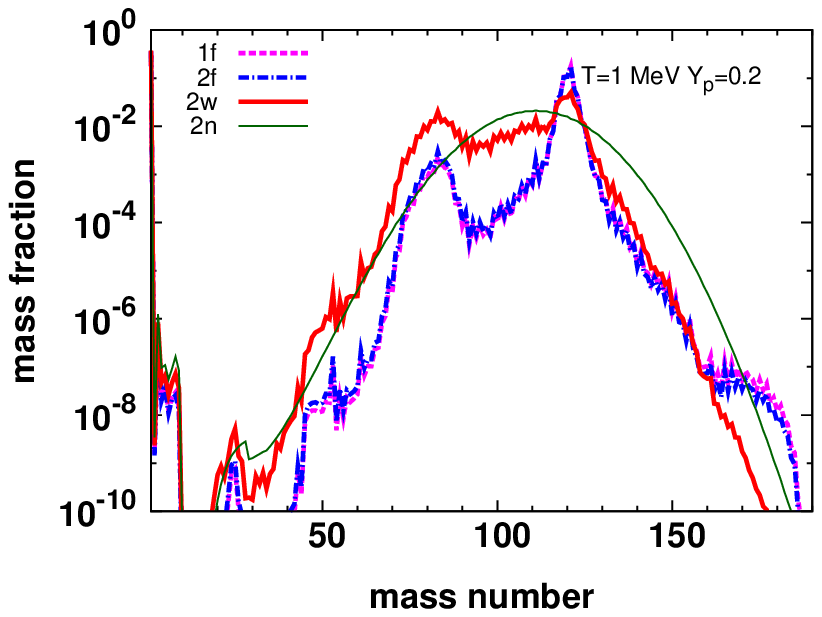}
\includegraphics[width=8cm]{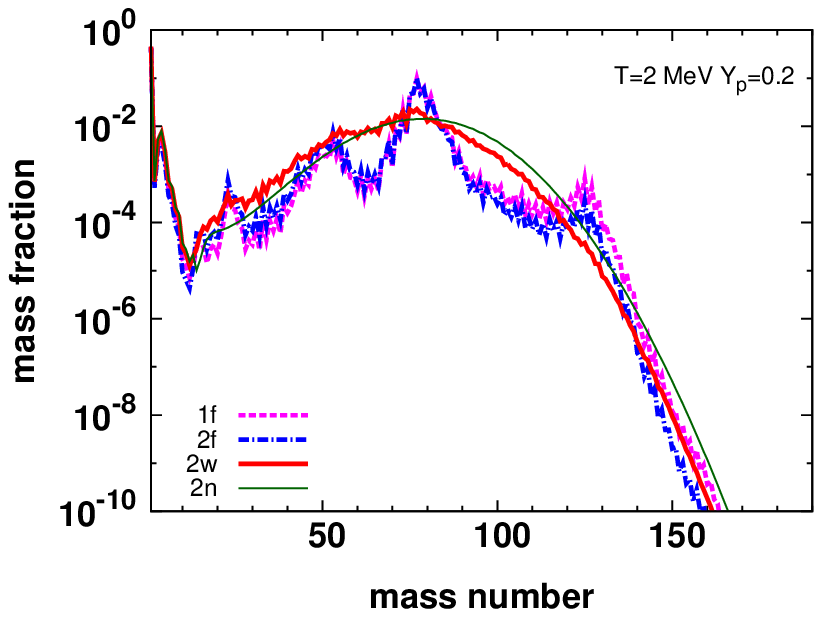}
\includegraphics[width=8cm]{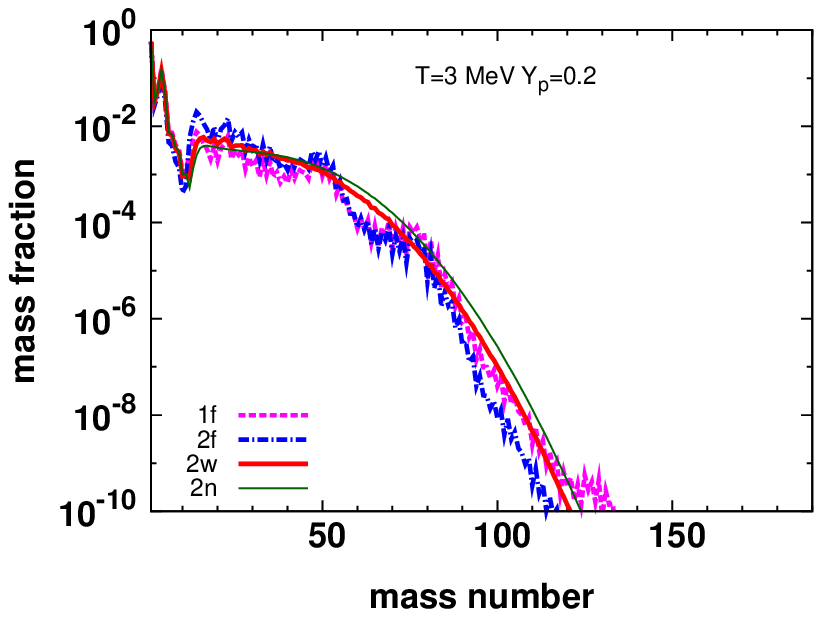}
\ \\
\ \\
\ \\
\caption{The 
mass fractions of elements as a function of  the mass number
 for Models~1f (magenta dashed lines), 2f  (blue dash dotted  lines), 2w  (red solid thick lines), 2n  (green solid thin lines) at  $T=$ 0.5 (left top panel), 1.0 (right top panel), 2.0 (left bottom panel)  and 3.0  (right bottom panel) MeV, $\rho_B=10^{12} $ g/cm$^3$  and $Y_p=0.2$. }
\label{fig_massdis02}
\end{figure}

\begin{figure}
\includegraphics[width=8cm]{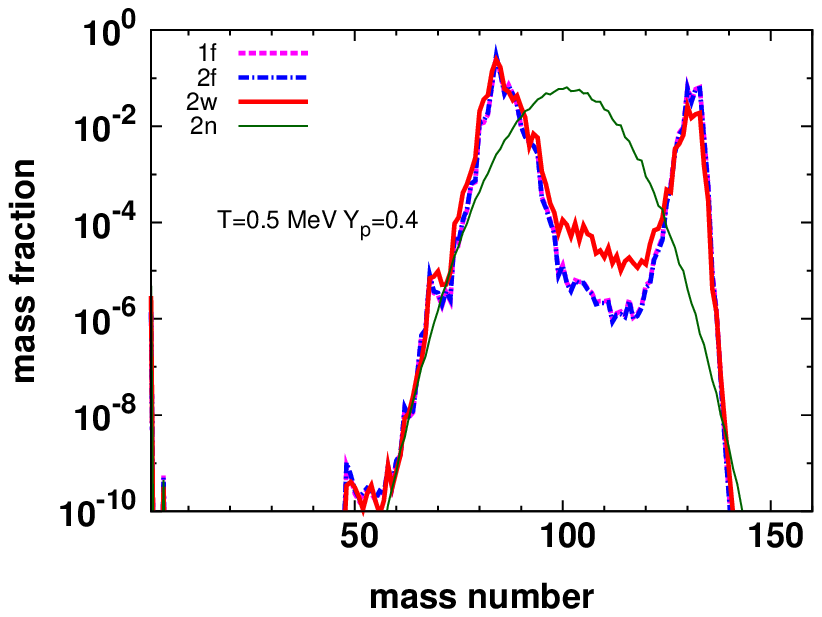}
\includegraphics[width=8cm]{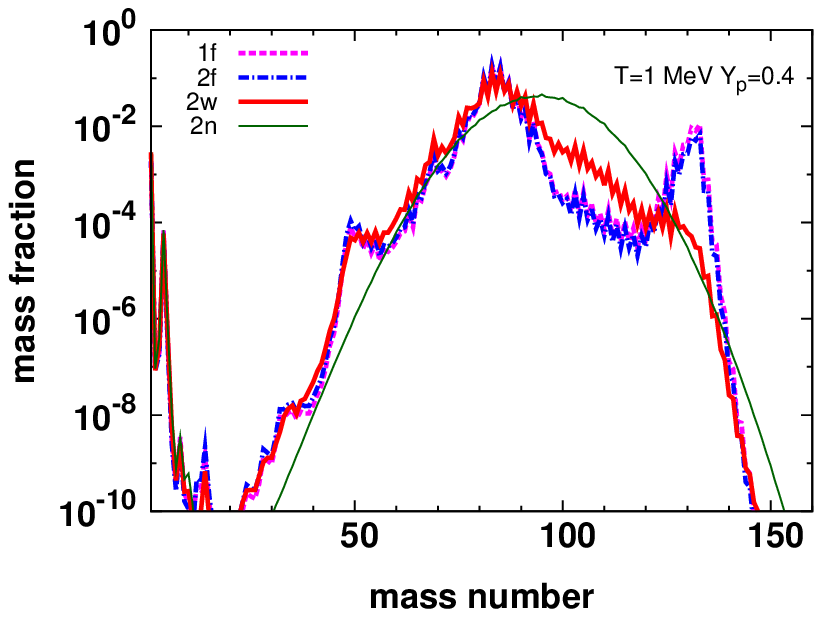}
\includegraphics[width=8cm]{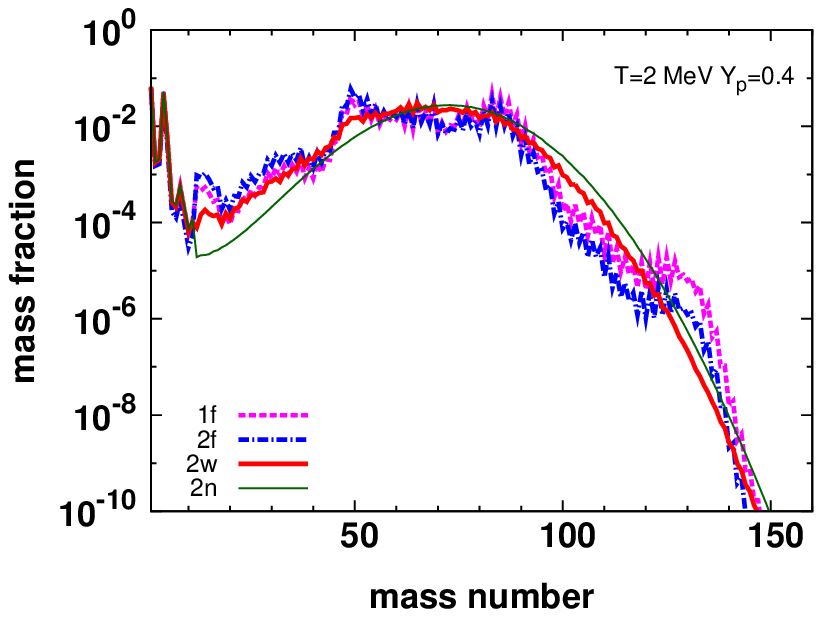}
\includegraphics[width=8cm]{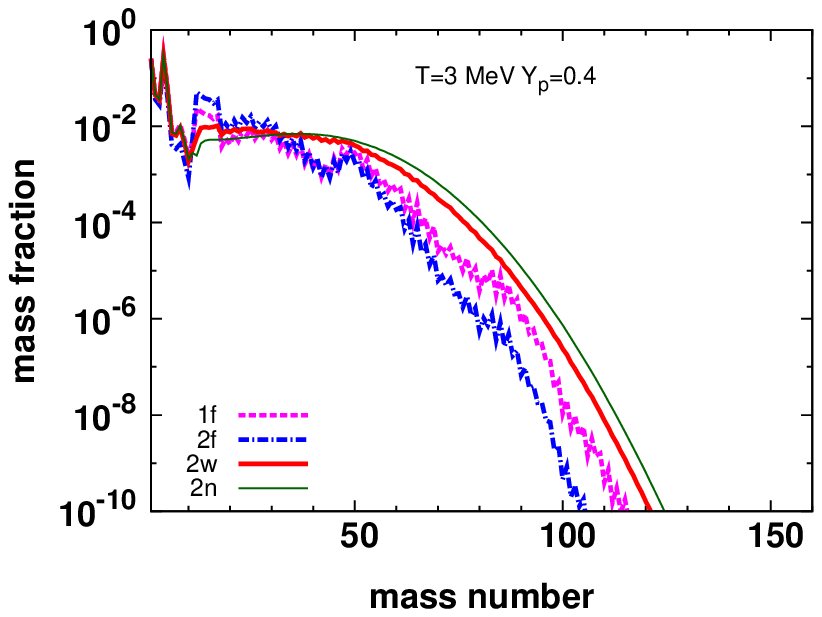}
\ \\
\ \\
\ \\
\caption{The 
mass fractions of elements as a function of  the mass number
for Models~1f (magenta dashed lines), 2f  (blue dash dotted  lines), 2w  (red solid thick lines), 2n  (green solid thin lines) at  $T=$ 0.5 (left top panel), 1.0 (right top panel), 2.0 (left bottom panel)  and 3.0  (right bottom panel) MeV, $\rho_B=10^{12} $ g/cm$^3$  and $Y_p=0.4$. }
\label{fig_massdis04}
\end{figure}

Figures~\ref{fig_massdis02}  and~\ref{fig_massdis04}  show the 
mass fractions of elements as a  function of  the mass number
 for Models 1f, 2f, 2w and 2n at $\rho_B=$~10$^{12}$~g/cm$^3$, $T=$~0.5, 1.0, 2.0 and 3.0 MeV  and  $Y_p=$~0.2 and 0.4. 
We can see from the comparison of  Models 1f and 2f  that the temperature dependence in the surface energies has no influence on the nuclear abundance below $T=1$~MeV.
At  $T=2$ and 3 MeV, 
the population of small-mass number elements is larger in Model 2f than 
in Model 1f 
 due to the reduction of surface tension. 
The temperature dependence in the shell effects affects more on the element distribution at low temperatures  than that in the surface tension. 
It can be seen in Model 2w that the washout effect  makes the element distributions smoother compared to
the case with no washout
 (Model 2f). 
The differences are visible even at $T=$~0.5~MeV both for $Y_p=$~0.2 and 0.4.
We find that this effect reduces the mass fractions
 of nuclei in the vicinity of the neutron magic numbers 
 $N =28, 50, 82$ and $126$, which correspond to the peaks at $A \sim 50, 80, 120$ and $170$ in the figures, respectively.
It 
is evident that the peak at a
larger mass number is more strongly suppressed
since  the energy spacing, $\epsilon_i^{sh}$, is smaller for nuclei with larger mass numbers.
For instance, the third peak  disappears for  $T=$~1 MeV and  $Y_p=$~0.4.
For $T=$~2 and 3 MeV,
 the shell effects are almost gone
and 
there is no discernible feature remaining at the magic numbers in Model 2w 
 as in Model 2n, in which no shell effect is taken into account.
%
%

\begin{figure}
\includegraphics[width=8cm]{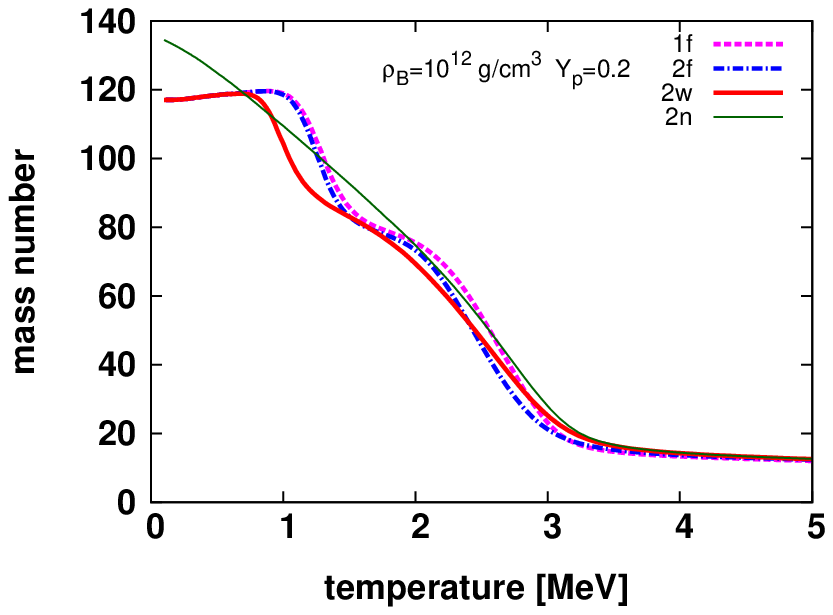}
\includegraphics[width=8cm]{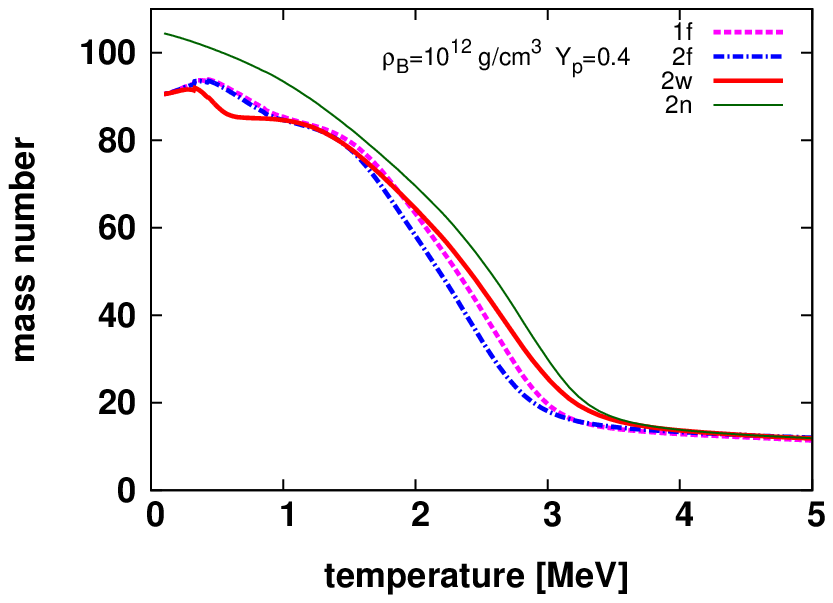}
\includegraphics[width=8cm]{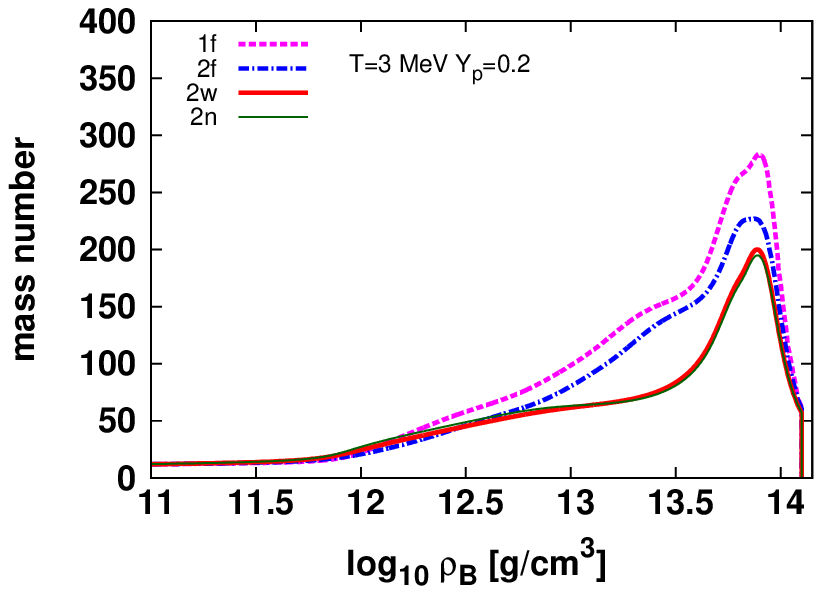}
\includegraphics[width=8cm]{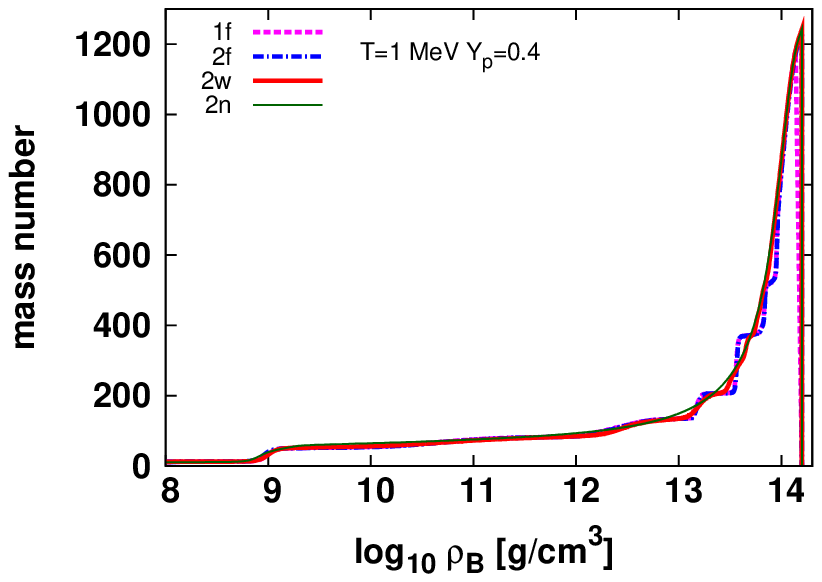}
\ \\
\ \\
\ \\
\caption{
 Average mass number of heavy nuclei with $Z \geq 6 $ for  for Models~1f (magenta dashed lines), 2f  (blue dash dotted  lines), 2w  (red solid thick lines), 2n  (green solid thin lines)
 at  $Y_p=$ 0.2 (left top panel) and 0.4 (right top panel) and  $\rho_B=10^{12} $ g/cm$^3$ as a function of temperature  as well as at  $T=$ 3~MeV and     $Y_p=$ 0.2 (left bottom panel) and   $T=$ 1~MeV  and  $Y_p=$ 0.4 (right bottom panel)  as a function of density.
}
\label{fig_massnumber}
\end{figure}

The average mass number of heavy nuclei ($Z_i \geq 6$) is displayed in Fig.~\ref{fig_massnumber} 
 at $\rho_B=$~10$^{12}$~g/cm$^3$  and  $Y_p=$0.2 and 0.4 as a function of  temperature 
and at  $T$=~3 MeV, $Y_p=$0.2 and
  $T$=~1 MeV, $Y_p=$0.4 as a function of density.
The difference between Models 1f and 2f  indicates that the temperature dependence in the surface tension cuts down the mass number of nuclei  above $T\sim2$~MeV,
which can be confirmed also in Figs.~\ref{fig_massdis02}~and~\ref{fig_massdis04}. 
The washout of shell effects tends to reduce the averaged mass number around $T \sim 1$~MeV for  $\rho=$~10$^{12}$~g/cm$^3$  and  $Y_p=$0.2 and 0.4  (top panels).
This is due to the reduction  of  the third peak at $N=82$ as explained in the previous paragraph.
Models 2w and 2n become almost identical around $T \sim $~3~MeV, which means the shell effects are completely extinct. 
At high densities for $T=$~3 MeV and $Y_p$=~0.2,  we can see  the average mass number in Model 2w  is decreased by the temperature effects both in the  surface and shell energies.
Note that the decline of the average mass number around $\rho\sim 10^{14}$~g/cm$^3$ 
for  $T=$~3 MeV and $Y_p$=~0.2 is due to  the formation of nuclear pastas 
 as discussed in detail in the previous paper 
\citep{furusawa11}.
It is interesting that for $T=1~$MeV and $Y_p$=0.4,
Model 2w is essentially identical to
Model  2f at low densities but 
gets closer to Model 2n at  $\rho \gtrsim 10^{13.5}$~g/cm$^3$.
This is because the shell effects of nuclei with larger mass numbers are more likely to be washed out  as already explained.
It is  also remarkable that the average mass number in Model 2w does not always settle down to a value 
between those in  Models 2f and 2n, since the shell effects 
are sensitive to the nuclear species and the washout affects the average mass number nonlinearly.  

\begin{figure}
\includegraphics[width=8cm]{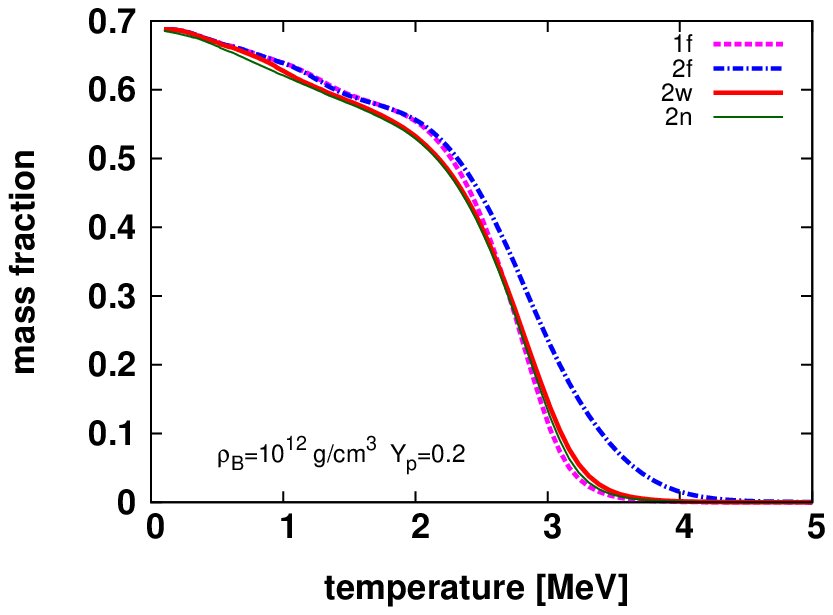}
\includegraphics[width=8cm]{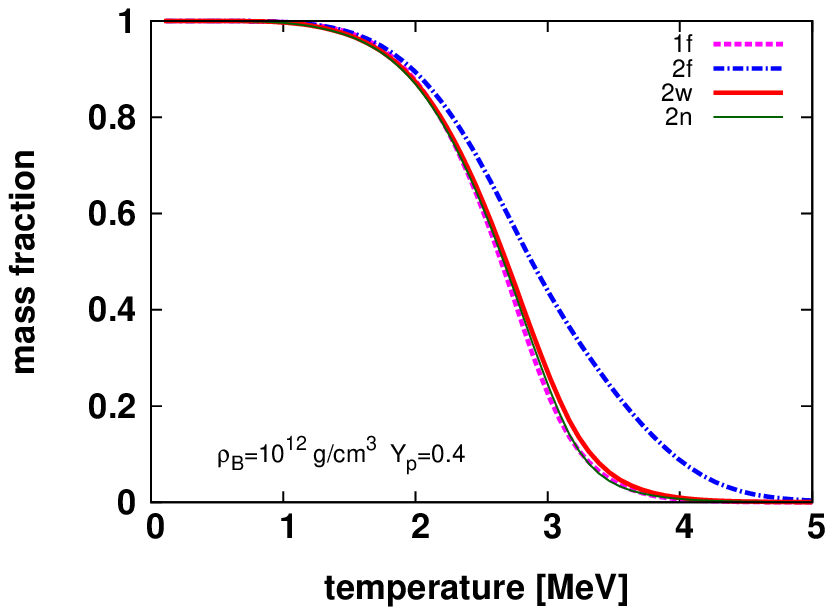}
\includegraphics[width=8cm]{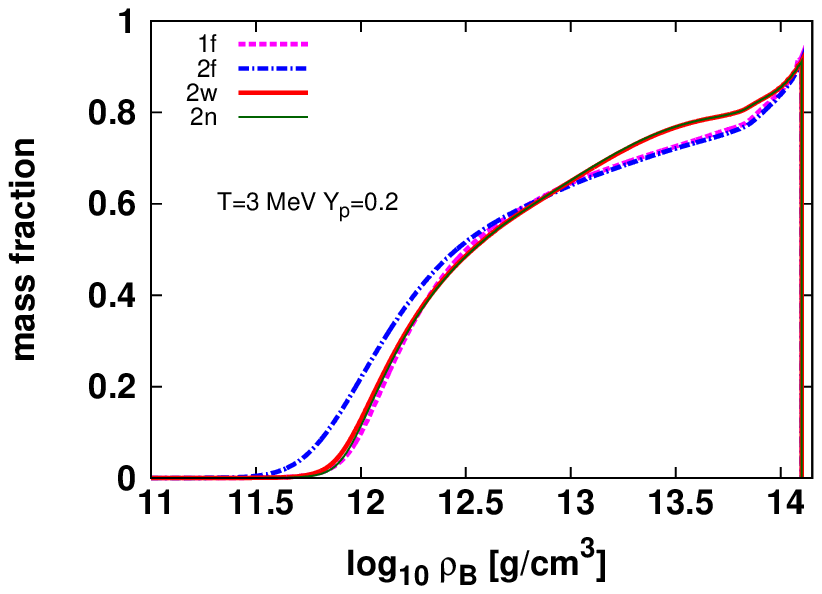}
\includegraphics[width=8cm]{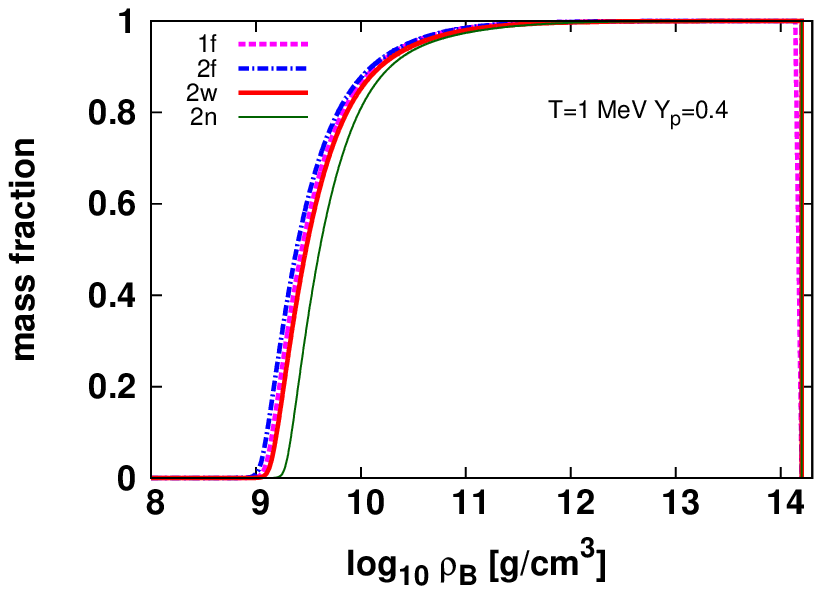}
\ \\
\ \\
\ \\
\caption{Mass fractions of heavy nuclei with $Z \geq 6 $ for  for Models~1f (magenta dashed lines), 2f  (blue dash dotted  lines), 2w  (red solid thick lines), 2n  (green solid thin lines)
 at  $Y_p=$ 0.2 (left top panel) and 0.4 (right top panel) and  $\rho_B=10^{12} $ g/cm$^3$ as a function of temperature  as well as at  $T=$ 3~MeV and     $Y_p=$0.2 (left bottom panel) and   $T=$ 1~MeV  and  $Y_p=$0.4 (right bottom panel)  as a function of density.
}
\label{fig_massfrac}
\end{figure}

The total mass fractions of the heavy nuclei are  shown in Fig.~\ref{fig_massfrac} 
at $\rho_B=$~10$^{12}$~g/cm$^3$  and  $Y_p=$0.2 and 0.4 as a function of  temperature 
and at  $T$=~3 MeV, $Y_p=$0.2 and
  $T$=~1 MeV, $Y_p=$0.4 as a function of density.
The mass fraction in Model 2f is larger than that in Model 1f at high temperatures, since the reduction of surface tension increases the binding energies  (or decreases the free energies) of heavy nuclei. On the other hand, the washout of shell effect in Model 2w cuts down their mass fraction for $\rho_B=$~10$^{12}$~g/cm$^3$ and  $Y_p=$0.2 and 0.4  (upper panels). 
The densities, where heavy nuclei emerge, are also affected by the difference in the shell effects and are lower in the order of Models 2f, 2w, 2n as shown  in the bottom panels.

Figure~\ref{fig_pre} shows the absolute values of  baryonic pressure for all models
at $\rho_B=$~10$^{12}$~g/cm$^3$  and  $Y_p=$0.2 and 0.4 as a function of  temperature 
and at  $T$=~3 MeV, $Y_p=$0.2 and
  $T$=~1 MeV, $Y_p=$0.4 as a function of density.
At low  temperatures and high densities, the pressures is negative, since the Coulomb energies of heavy nuclei  decreases by the compression 
\citep{furusawa11,furusawa13a}. 
We can see that the pressure is hardly affected
either
 by the modifications for the light nuclei other than $d$, $t$, $h$ and $\alpha$
 or by
the introductions of  the temperature dependences
in the surface tension and shell effect. 
It is also the case for the
 entropy per baryon as shown in Fig.~\ref{fig_ent}.
It is slightly affected, however, by the difference in the mass fractions of nucleons and nuclei
as displayed in Figs.~\ref{fig_masslight} and \ref{fig_massfrac}.

\section{Summary and Discussion}
We have constructed the improved baryonic EOS with multi-nucleus for the use in the simulation of core-collapse supernovae. This EOS provides the abundance of various nuclei with
 various in-medium effects being taken into consideration at high densities and/or temperatures. The major improvement
in the new EOS is the introduction of the washout of shell effects 
 based on the single particle motion of nucleons outside the closed shell,
 which has been  ignored in any EOS employed so far for supernova simulations.  
We have also added the temperature dependence in the surface tensions
and replaced the mass formula by the liquid drop model for the  light nuclei ($Z\leq5$) other than $d$,  $t$, $h$ and $\alpha$ 
 with the  one based on the  quasi-particle description.

The basic part of the the model is the same as that given in \citet{furusawa13a}.  
The model free energy density is constructed so that it should reproduce the ordinary NSE results at low densities and temperatures  and make a continuous transition to the supra-nuclear density EOS obtained from the RMF with the TM1 parameter set. 
The free energy density of the nucleon vapor outside nuclei is also calculated by the same RMF.  
The contribution from heavy nuclei is evaluated with the original LDM, in which their bulk energies are obtained again via
the same RMF
 and various in-medium effects are taken
 into account for the surface and Coulomb energies such as the existence of the pasta phase near the nuclear saturation density.
 The shell energies
in the vacuum limit
  are  obtained from  either experimental or theoretical data.
The light nuclei $d$,  $t$, $h$ and $\alpha$ are handled with
 the quantum approach, in which the self- and Pauli-energy shifts are taken into account. 

 We have compared the abundances of nuclei 
as well as the thermodynamical quantities obtained in different models for some combinations of density, temperature and proton fraction.
In the new model, the mass fraction of  the light nuclei other than  $d$,  $t$, $h$ and $\alpha$ is reduced  a little  due to the Pauli energy shifts. 
The temperature dependence in the surface energies for heavy nuclei 
leads solely to smaller average mass numbers and larger fractions of them  above $T \sim 2$ MeV than before.
This is because the reduction of surface energies requires smaller
Coulomb energies.
We have found that the introduction of the washout of shell effects greatly changes
the compositions.
The shell effects are washed out slightly  even at $T \sim 1$ MeV and  almost disappear
at $T \sim 2-3$ MeV. 
 These results are consistent with 
the findings by  \cite{nishimura14}.
It has been also shown that the shell effects for nuclei with larger mass numbers are more strongly washed out
because of the smaller energy spacing among the shells.
It has been found that the modifications to the  model free energy in this paper 
have little influence on thermodynamical quantities such as pressure and entropy. 

The model free energy in the new EOS still has room for further improvement,
 since there are many uncertainties in our phenomenological treatment of the in-medium effects.
We may need to sophisticate the EOS for uniform nuclear matter, on which  the evaluations of the free nucleons as well as bulk energies of heavy nuclei are based.
In fact, the RMF with TM1 parameter set  is known to have the symmetry energy of $J=$36.9 MeV which is  larger than the canonical value $29 \lesssim J  \lesssim 35$ MeV \citep{tews13}. 
It is noted, however, that  the TM1 set is known to reproduce  excellent
properties of the ground states of  stable nuclei as well as unstable nuclei and 
to agree well with experimental data in studies of nuclei with deformed configuration and the giant resonances within the RPA formalism \citep{sugahara94, hirata95, ma97,ma01}.
The employment of
another EOS for uniform nuclear matter \citep{togashi13} is  indeed under progress. 
We would like to stress, however, that the new EOS
constructed in this paper is
more realistic than the previous one
and that the changes in the element distribution made
 by the washout of shell effects may have a great impact on the
 weak interaction rate in the collapsing cores of massive stars. 
The systematical study of  such effects
 is  also underway. 

\section*{Acknowledgments}%
S.F. gratefully acknowledges S.~Nishimura, M.~Takano, M.~Hempel and I.~Mishustin for
their useful discussions. S.F. is supported by Japan Society for the Promotion of
Science Postdoctoral Fellowships for Research Abroad.
Some numerical calculations were carried out on  PC cluster at Center
for Computational Astrophysics, National Astronomical Observatory of Japan.
This work is supported in part by the usage of supercomputer systems
through the Large Scale Simulation Program
(No. 15/16/-08) of High Energy Accelerator Research Organization (KEK)
and
Post-K Projects  (hp160071, hp160211) at K-computer, RIKEN AICS
as well as the computational resources provided by
RCNP at Osaka University, YITP at Kyoto University, University of Tokyo
and JLDG.
This work was
supported by Grant-in-Aid for the Scientiﬁc Research
from the Ministry of Education, Culture, Sports, Science
and Technology (MEXT), Japan (24103006, 24244036,16H03986,15K05093, 24105008, 26104006).

\section*{Appendix: Table data}
Our EOS tables  are available on the Web at
http://user.numazu-ct.ac.jp/~sumi/eos/.
Below we explain the detailed information.
The grid points and spacing  for $T$, $Y_p$ and $\rho_B$ are just the  same
 as those of \cite{shen11}, which is summarized in Table~\ref{tab_table}.
The results in the order of increasing $Y_p$ and $\rho_B$ are listed for each $T$.
The temperature and its logarithm, log$_{10}(T)$,  in units of  MeV  are written at the beginning of the each block.
The quantities in one line 
defined as follows:
\begin{enumerate}
\item Logarithm of baryon mass density:  log$_{10}(\rho_B)$  [g/cm$^3$]  ($\rho_B=n_B m_u$)
\item Baryon number density: $n_B$ [fm$^{-3}$]
\item Proton Fraction: $Y_p$
\item Free energy: $F=f/n_B-M_B$ [MeV]
\item Internal energy per baryon: $E_{int}=f/n_B-sT -m_u$ [MeV]
\item Entropy per baryon: $s$  [k$_B$]
\item Average mass number of  the heavy nuclei ($Z\geq6$): $ <A_i >$
\item Average proton number of the heavy nuclei ($Z\geq6$): $<Z_i>$
\item Effective mass: $m^{\ast}$ [MeV]
\item Free neutron fraction: $X_n$
\item Free proton fraction: $X_p$
\item  Fraction of  the light nuclei ($Z\leq5$): $X_a$
\item Fraction of  the heavy nuclei  ($Z\geq6$): $X_A$
\item Baryon Pressure: $p_B$[MeV/fm$^{-3}$]
\item Chemical potential of the neutron: $ \mu_n =  (\partial{f}/\partial{\tilde{n}_n}  )_{\tilde{n}_p} - m_B $[MeV]  \\ ($\tilde{n_n}=(1-Y_p)  n_B$ and $\tilde{n_p}= Y_p n_B $)
\item Chemical potential of the proton:  $\mu_p=  (\partial{f}/\partial{\tilde{n}_p} )_{\tilde{n}_n} - m_B $[MeV]
\item Average mass number of  the light nuclei ($Z\leq5$): $ <A_j >$
\item Average proton number of the light nuclei ($Z\leq5$): $ <Z_j >$
\item Deuteron fraction: $X_d$
\item Triton fraction: $X_t$
\item Helion fraction: $X_h$
\item Alpha-particle fraction: $X_{\alpha}$
\item Fraction of  the light nuclei  other than $d$, $t$, $h$ and $\alpha$ ($Z\leq5$): $X_{a'}$ ($X_a=X_{a'}+X_d+X_t+X_h+X_{\alpha}$)
\item Average mass number of  the light nuclei  other than $d$, $t$, $h$ and $\alpha$ ($Z\leq5$): $<A_{j'}>$
\item Average proton number of  the light nuclei  other than $d$, $t$, $h$ and $\alpha$  ($Z\leq5$): $<Z_{j'}>$
\end{enumerate}

\begin{table}[t]
\begin{tabular}{|c||c|c|c||c|}
\hline
   model & light nuclei  & $T$ dependence in  & shell energies & \ version \\
   & other than  $d,t,h,\alpha$ & surface energies & &  \\
 \hline
 \hline
 0f &  LDM & no  & full    & FYSS13 \\
 1f & quasi-particle &  no & full    &  \\
 2f &  quasi-particle  & yes & full     & \\
 2w & quasi-particle   & yes & washout     &  FYSS16 \\
 2n &  quasi-particle   & yes & no    & \\
\hline
\end{tabular}
\caption{\label{tab1_model}%
Different models for comparisons. See the beginning of  Sec.~\ref{results}. }  
\end{table}

\begin{table}[t]
\begin{tabular}{|c||c|c|c|}
\hline
    & Range  &  Grid spacing  & Points\\
 \hline
 \hline
$T$         &    $-1.0\leq$log$_{10}(T)\leq$2.6   & $\Delta$log$_{10}(T)$=0.04      & 91 \\
$Y_p$      &  0.01$\leq Y_p \leq$0.65     &  0.01    & 65 \\
$\rho_B$  & 5.1$\leq $log$_{10}\rho_B \leq$ 16.0   &   $\Delta$log$_{10}\rho_B =0.1$    & 110 \\
\hline
\end{tabular}
\caption{\label{tab_table}%
The EOS table of Model 2w (the same format as that of \cite{shen11}).}  
\end{table}


  \bibliographystyle{elsarticle-harv} 
  \bibliography{reference160724}

\begin{thebibliography}{43}
\expandafter\ifx\csname natexlab\endcsname\relax\def\natexlab#1{#1}\fi
\expandafter\ifx\csname url\endcsname\relax
  \def\url#1{\texttt{#1}}\fi
\expandafter\ifx\csname urlprefix\endcsname\relax\def\urlprefix{URL }\fi

\bibitem[{{Agrawal} et~al.(2014){Agrawal}, {De}, {Samaddar}, {Centelles}, and
  {Vi{\~n}as}}]{agrawal14a}
{Agrawal}, B.~K., {De}, J.~N., {Samaddar}, S.~K., {Centelles}, M., {Vi{\~n}as},
  X., Feb. 2014. {Symmetry energy of warm nuclear systems}. European Physical
  Journal A 50, 19.

\bibitem[{{Audi} et~al.(2012){Audi}, {M.}, {A.~H.}, {F.~G.}, {MacCormick},
  {Xu}, and {Pfeiffer}}]{audi12}
{Audi}, G., {M.}, W., {A.~H.}, W., {F.~G.}, K., {MacCormick}, M., {Xu}, X.,
  {Pfeiffer}, B., Dec. 2012. {The Ame2012 atomic mass evaluation}. Chinese
  Physics C 36, 002.

\bibitem[{{Aymard} et~al.(2014){Aymard}, {Gulminelli}, and
  {Margueron}}]{aymard14}
{Aymard}, F., {Gulminelli}, F., {Margueron}, J., Jun. 2014. {In-medium nuclear
  cluster energies within the extended Thomas-Fermi approach}. \prc 89~(6),
  065807.

\bibitem[{{Blinnikov} et~al.(2011){Blinnikov}, {Panov}, {Rudzsky}, and
  {Sumiyoshi}}]{blinnikov11}
{Blinnikov}, S.~I., {Panov}, I.~V., {Rudzsky}, M.~A., {Sumiyoshi}, K., Nov.
  2011. {The equation of state and composition of hot, dense matter in
  core-collapse supernovae}. \aap 535, A37.

\bibitem[{Bohr and Mottelson(1998)}]{bohr87}
Bohr, A., Mottelson, B., 1998. Nuclear Structure. No. v. 2 in Nuclear
  Structure. World Scientific.
\newline\urlprefix\url{https://books.google.com.au/books?id=l1pshSMQwcYC}

\bibitem[{{Bondorf} et~al.(1995){Bondorf}, {Botvina}, {Iljinov}, {Mishustin},
  and {Sneppen}}]{bondorf95}
{Bondorf}, J.~P., {Botvina}, A.~S., {Iljinov}, A.~S., {Mishustin}, I.~N.,
  {Sneppen}, K., Jun. 1995. {Statistical multifragmentation of nuclei}.
  \physrep 257, 133--221.

\bibitem[{{Botvina} and {Mishustin}(2004)}]{botvina04}
{Botvina}, A.~S., {Mishustin}, I.~N., Apr. 2004. {Formation of hot heavy nuclei
  in supernova explosions}. Physics Letters B 584, 233--240.

\bibitem[{{Botvina} and {Mishustin}(2010)}]{botvina10}
{Botvina}, A.~S., {Mishustin}, I.~N., Oct. 2010. {Statistical approach for
  supernova matter}. Nuclear Physics A 843, 98--132.

\bibitem[{{Brack} and {Quentin}(1974)}]{brack74}
{Brack}, M., {Quentin}, P., Sep. 1974. {Selfconsistent calculations of highly
  excited nuclei}. Physics Letters B 52, 159--162.

\bibitem[{{Burrows}(2013)}]{burrows13}
{Burrows}, A., Jan. 2013. {Colloquium: Perspectives on core-collapse supernova
  theory}. Reviews of Modern Physics 85, 245--261.

\bibitem[{{Burrows} and {Lattimer}(1984)}]{burrows84}
{Burrows}, A., {Lattimer}, J.~M., Oct. 1984. {On the accuracy of the
  single-nucleus approximation in the equation of state of hot, dense matter}.
  \apj 285, 294--303.

\bibitem[{{Buyukcizmeci} et~al.(2014){Buyukcizmeci}, {Botvina}, and
  {Mishustin}}]{buyukcizmeci14}
{Buyukcizmeci}, N., {Botvina}, A.~S., {Mishustin}, I.~N., Jul. 2014. {Tabulated
  Equation of State for Supernova Matter Including Full Nuclear Ensemble}. \apj
  789, 33.

\bibitem[{{Buyukcizmeci} et~al.(2013){Buyukcizmeci}, {Botvina}, {Mishustin},
  {Ogul}, {Hempel}, {Schaffner-Bielich}, {Thielemann}, {Furusawa}, {Sumiyoshi},
  {Yamada}, and {Suzuki}}]{buyukcizmeci13}
{Buyukcizmeci}, N., {Botvina}, A.~S., {Mishustin}, I.~N., {Ogul}, R., {Hempel},
  M., {Schaffner-Bielich}, J., {Thielemann}, F.-K., {Furusawa}, S.,
  {Sumiyoshi}, K., {Yamada}, S., {Suzuki}, H., Jun. 2013. {A comparative study
  of statistical models for nuclear equation of state of stellar matter}.
  Nuclear Physics A 907, 13--54.

\bibitem[{{Furusawa} et~al.(2013){Furusawa}, {Sumiyoshi}, {Yamada}, and
  {Suzuki}}]{furusawa13a}
{Furusawa}, S., {Sumiyoshi}, K., {Yamada}, S., {Suzuki}, H., Aug. 2013. {New
  Equations of State Based on the Liquid Drop Model of Heavy Nuclei and Quantum
  Approach to Light Nuclei for Core-collapse Supernova Simulations}. \apj 772,
  95.

\bibitem[{{Furusawa} et~al.(2011){Furusawa}, {Yamada}, {Sumiyoshi}, and
  {Suzuki}}]{furusawa11}
{Furusawa}, S., {Yamada}, S., {Sumiyoshi}, K., {Suzuki}, H., Sep. 2011. {A New
  Baryonic Equation of State at Sub-nuclear Densities for Core-collapse
  Simulations}. \apj 738, 178.

\bibitem[{{Hempel} and {Schaffner-Bielich}(2010)}]{hempel10}
{Hempel}, M., {Schaffner-Bielich}, J., Jun. 2010. {A statistical model for a
  complete supernova equation of state}. Nuclear Physics A 837, 210--254.

\bibitem[{{Hirata} et~al.(1995){Hirata}, {Toki}, and {Tanihata}}]{hirata95}
{Hirata}, D., {Toki}, H., {Tanihata}, I., Feb. 1995. {Relativistic mean-field
  theory on the xenon, cesium and barium isotopes}. Nuclear Physics A 589,
  239--248.

\bibitem[{{Hix} et~al.(2003){Hix}, {Messer}, {Mezzacappa}, {Liebend{\"o}rfer},
  {Sampaio}, {Langanke}, {Dean}, and {Mart{\'{\i}}nez-Pinedo}}]{hix03}
{Hix}, W.~R., {Messer}, O.~E., {Mezzacappa}, A., {Liebend{\"o}rfer}, M.,
  {Sampaio}, J., {Langanke}, K., {Dean}, D.~J., {Mart{\'{\i}}nez-Pinedo}, G.,
  Nov. 2003. {Consequences of Nuclear Electron Capture in Core Collapse
  Supernovae}. Physical Review Letters 91~(20), 201102.

\bibitem[{{Janka}(2012)}]{janka12}
{Janka}, H.-T., Nov. 2012. {Explosion Mechanisms of Core-Collapse Supernovae}.
  Annual Review of Nuclear and Particle Science 62, 407--451.

\bibitem[{{Kotake} et~al.(2012){Kotake}, {Takiwaki}, {Suwa}, {Iwakami Nakano},
  {Kawagoe}, {Masada}, and {Fujimoto}}]{kotake12}
{Kotake}, K., {Takiwaki}, T., {Suwa}, Y., {Iwakami Nakano}, W., {Kawagoe}, S.,
  {Masada}, Y., {Fujimoto}, S.-i., 2012. {Multimessengers from Core-Collapse
  Supernovae: Multidimensionality as a Key to Bridge Theory and Observation}.
  Advances in Astronomy 2012, 428757.

\bibitem[{{Koura} et~al.(2005){Koura}, {Tachibana}, {Uno}, and
  {Yamada}}]{koura05}
{Koura}, H., {Tachibana}, T., {Uno}, M., {Yamada}, M., Feb. 2005. {Nuclidic
  Mass Formula on a Spherical Basis with an Improved Even-Odd Term}. Progress
  of Theoretical Physics 113, 305--325.

\bibitem[{{Lattimer} and {Swesty}(1991)}]{lattimer91}
{Lattimer}, J.~M., {Swesty}, F.~D., Dec. 1991. {A generalized equation of state
  for hot, dense matter}. Nuclear Physics A 535, 331--376.

\bibitem[{{Lentz} et~al.(2012){Lentz}, {Mezzacappa}, {Messer}, {Hix}, and
  {Bruenn}}]{lentz12}
{Lentz}, E.~J., {Mezzacappa}, A., {Messer}, O.~E.~B., {Hix}, W.~R., {Bruenn},
  S.~W., Nov. 2012. {Interplay of Neutrino Opacities in Core-collapse Supernova
  Simulations}. \apj 760, 94.

\bibitem[{{Ma} et~al.(1997){Ma}, {Toki}, {Chen}, and {Giai}}]{ma97}
{Ma}, Z., {Toki}, H., {Chen}, B., {Giai}, N.~V., Oct. 1997. {The Giant Dipole
  Resonance in Ar-Isotopes in the Relativistic RPA}. Progress of Theoretical
  Physics 98, 917--926.

\bibitem[{{Ma} et~al.(2001){Ma}, {Van Giai}, {Wandelt}, {Vretenar}, and
  {Ring}}]{ma01}
{Ma}, Z.-y., {Van Giai}, N., {Wandelt}, A., {Vretenar}, D., {Ring}, P., Apr.
  2001. {Isoscalar compression modes in relativistic random phase
  approximation}. Nuclear Physics A 686, 173--186.

\bibitem[{{Newton} and {Stone}(2009)}]{newton09}
{Newton}, W.~G., {Stone}, J.~R., May 2009. {Modeling nuclear ``pasta'' and the
  transition to uniform nuclear matter with the 3D Skyrme-Hartree-Fock method
  at finite temperature: Core-collapse supernovae}. \prc 79~(5), 055801.

\bibitem[{{Nishimura} and {Takano}(2014)}]{nishimura14}
{Nishimura}, S., {Takano}, M., May 2014. {Shell effects in hot nuclei and their
  influence on nuclear composition in supernova matter}. In: {Jeong}, S.,
  {Imai}, N., {Miyatake}, H., {Kajino}, T. (Eds.), American Institute of
  Physics Conference Series. Vol. 1594 of American Institute of Physics
  Conference Series. pp. 239--244.

\bibitem[{{Okamoto} et~al.(2012){Okamoto}, {Maruyama}, {Yabana}, and
  {Tatsumi}}]{okamoto12}
{Okamoto}, M., {Maruyama}, T., {Yabana}, K., {Tatsumi}, T., Jul. 2012.
  {Three-dimensional structure of low-density nuclear matter}. Physics Letters
  B 713, 284--288.

\bibitem[{{Raduta} et~al.(2016){Raduta}, {Gulminelli}, and {Oertel}}]{raduta16}
{Raduta}, A.~R., {Gulminelli}, F., {Oertel}, M., Feb. 2016. {Modification of
  magicity toward the dripline and its impact on electron-capture rates for
  stellar core collapse}. \prc 93~(2), 025803.

\bibitem[{{Ravenhall} et~al.(1983){Ravenhall}, {Pethick}, and
  {Wilson}}]{ravenhall83}
{Ravenhall}, D.~G., {Pethick}, C.~J., {Wilson}, J.~R., Jun. 1983. {Structure of
  Matter below Nuclear Saturation Density}. Physical Review Letters 50,
  2066--2069.

\bibitem[{{R{\"o}pke}(2009)}]{roepke09}
{R{\"o}pke}, G., Jan. 2009. {Light nuclei quasiparticle energy shifts in hot
  and dense nuclear matter}. \prc 79~(1), 014002.

\bibitem[{{Sandulescu} et~al.(1997){Sandulescu}, {Civitarese}, {Liotta}, and
  {Vertse}}]{sandulescu97}
{Sandulescu}, N., {Civitarese}, O., {Liotta}, R.~J., {Vertse}, T., Mar. 1997.
  {Effects due to the continuum on shell corrections at finite temperatures}.
  \prc 55, 1250--1254.

\bibitem[{{Shen} et~al.(2011){Shen}, {Horowitz}, and {Teige}}]{sheng11}
{Shen}, G., {Horowitz}, C.~J., {Teige}, S., Mar. 2011. {New equation of state
  for astrophysical simulations}. \prc 83~(3), 035802.

\bibitem[{{Shen} et~al.(1998{\natexlab{a}}){Shen}, {Toki}, {Oyamatsu}, and
  {Sumiyoshi}}]{shen98a}
{Shen}, H., {Toki}, H., {Oyamatsu}, K., {Sumiyoshi}, K., Jul.
  1998{\natexlab{a}}. {Relativistic equation of state of nuclear matter for
  supernova and neutron star}. Nuclear Physics A 637, 435--450.

\bibitem[{{Shen} et~al.(1998{\natexlab{b}}){Shen}, {Toki}, {Oyamatsu}, and
  {Sumiyoshi}}]{shen98b}
{Shen}, H., {Toki}, H., {Oyamatsu}, K., {Sumiyoshi}, K., Nov.
  1998{\natexlab{b}}. {Relativistic Equation of State of Nuclear Matter for
  Supernova Explosion}. Progress of Theoretical Physics 100, 1013--1031.

\bibitem[{Shen et~al.(2011)Shen, Toki, Oyamatsu, and Sumiyoshi}]{shen11}
Shen, H., Toki, H., Oyamatsu, K., Sumiyoshi, K., 2011. {Relativistic Equation
  of State for Core-Collapse Supernova Simulations}. Astrophys.J.Suppl. 197,
  20.

\bibitem[{{Steiner} et~al.(2013){Steiner}, {Hempel}, and {Fischer}}]{steiner13}
{Steiner}, A.~W., {Hempel}, M., {Fischer}, T., Sep. 2013. {Core-collapse
  Supernova Equations of State Based on Neutron Star Observations}. \apj 774,
  17.

\bibitem[{Sugahara and Toki(1994)}]{sugahara94}
Sugahara, Y., Toki, H., 1994. {Relativistic mean field theory for unstable
  nuclei with nonlinear sigma and omega terms}. Nucl.Phys. A579, 557--572.

\bibitem[{Tews et~al.(2013)Tews, Kr\"uger, Hebeler, and Schwenk}]{tews13}
Tews, I., Kr\"uger, T., Hebeler, K., Schwenk, A., Jan 2013. Neutron matter at
  next-to-next-to-next-to-leading order in chiral effective field theory. Phys.
  Rev. Lett. 110, 032504.
\newline\urlprefix\url{http://link.aps.org/doi/10.1103/PhysRevLett.110.032504}

\bibitem[{{Timmes} and {Arnett}(1999)}]{timmes99}
{Timmes}, F.~X., {Arnett}, D., Nov. 1999. {The Accuracy, Consistency, and Speed
  of Five Equations of State for Stellar Hydrodynamics}. \apjs 125, 277--294.

\bibitem[{{Togashi} and {Takano}(2013)}]{togashi13}
{Togashi}, H., {Takano}, M., Mar. 2013. {Variational study for the equation of
  state of asymmetric nuclear matter at finite temperatures}. Nuclear Physics A
  902, 53--73.

\bibitem[{{Typel} et~al.(2010){Typel}, {R{\"o}pke}, {Kl{\"a}hn}, {Blaschke},
  and {Wolter}}]{typel10}
{Typel}, S., {R{\"o}pke}, G., {Kl{\"a}hn}, T., {Blaschke}, D., {Wolter}, H.~H.,
  Jan. 2010. {Composition and thermodynamics of nuclear matter with light
  clusters}. \prc 81~(1), 015803.

\bibitem[{{Watanabe} et~al.(2005){Watanabe}, {Maruyama}, {Sato}, {Yasuoka}, and
  {Ebisuzaki}}]{watanabe05}
{Watanabe}, G., {Maruyama}, T., {Sato}, K., {Yasuoka}, K., {Ebisuzaki}, T.,
  Jan. 2005. {Simulation of Transitions between ``Pasta'' Phases in Dense
  Matter}. Physical Review Letters 94~(3), 031101.

\end{thebibliography}
\end{document}